\documentclass[a4paper,12pt]{article}
\usepackage{amssymb,amsmath,amsthm,graphicx,ulem}

\pdfoutput=1

\usepackage{graphicx,subfigure}
\usepackage{epsfig}
\usepackage{amsmath}
\usepackage{amsfonts}
\usepackage{amssymb}
\usepackage[usenames]{color}
\usepackage[colorlinks,
            linkcolor=blue,
            anchorcolor=blue,
            citecolor=green
            ]{hyperref}
\usepackage[a4paper,left=2.cm,right=2.cm,top=2.5cm,bottom=2.5cm]{geometry}

\numberwithin{equation}{section}

\begin{document}
%%%%%%%%%%%%%%%%%%%%%%%%%%%%%%%%%%%%%%%%%%%%%%%%%%%%%%%%%%%%%%%%%%%%%%%%%%%%

\allowdisplaybreaks

\normalem

\title{ Deforming  black holes with even multipolar differential rotation boundary }

\vskip1cm

\author{ Hong-Bo Li, Tong-Tong Hu,
 Ben-Shen Song,    \\Shuo Sun,
 and  Yong-Qiang Wang\footnote{ yqwang@lzu.edu.cn, corresponding author}
\\ \\
   Research Center of Gravitation $\&$
 Institute of Theoretical Physics $\&$ \\
Key Laboratory for Magnetism and Magnetic of the Ministry of Education,\\ Lanzhou University, Lanzhou 730000, China
 \\
}

\maketitle

\begin{abstract}
Motivated by the novel asymptotically  global AdS$_4$ solutions with deforming horizon in [JHEP {\bf 1802}, 060 (2018)], we analyze
the boundary metric  with even multipolar differential rotation  and numerically construct a family of deforming solutions with quadrupolar differential rotation boundary,
including two classes of solutions: solitons and black holes.
In contrast to solutions with dipolar differential rotation boundary,  we find that even though  the norm of Killing vector  $\partial_t$  becomes spacelike for certain regions of polar angle $\theta$ when $\varepsilon>2$,
solitons and black holes with quadrupolar differential rotation still exist and  do not develop hair due to superradiance.
 Moreover, at the same temperature, the horizonal deformation of quadrupolar rotation is smaller than that of   dipolar rotation.
Furthermore,  we also study the entropy and  quasinormal modes of  the solutions, which have the analogous properties to that
of dipolar rotation.
\end{abstract}

\newpage
%%%%%%%%%%%%%%%%%%%%%%%%%%%%%%%%%%%%%%%%%%%%%%%%%%%%%%%%%%%%%%%%%%
%%%%%%%%%%%%%%%%%%%%%%%%%%%%%%%%%%%%%%%%%%%%%%%%%%%%%%%%%%%%%%%%%%

\tableofcontents

%%%%%%%%%%%%%%%%%%%%%%%%%%%%%%%%%%%%%%%%%%%%%%%%%%%%%%%%%%%%%%%%%
%%%%%%%%%%%%%%%%%%%%%%%%%%%%%%%%%%%%%%%%%%%%%%%%%%%%%%%%%%%%%%%%%
\section{Introduction}
According to uniqueness theorem of black holes \cite{Israel:1967jk,Ruffini67jk,Carterjk,Chrusciel:2012jk} in classical general relativity, the four-dimensional, asymptotically flat black hole solutions of zero angular momentum is identically  a family of Schwarzschild  black holes, whose
event horizon is a sphere surface. In four-dimensional anti-de Sitter (AdS) spacetime,
one found that except for compact
horizons of arbitrary genus,   there exists the black holes with noncompact planar or
negative constant curvature hyperbolic horizons.
It is  important to study physical properties and  applications of the asymptotically AdS black holes, especially, such black holes have recently been of great interest in the context of the Anti-de Sitter/conformal field theory (AdS/CFT) correspondence  \cite{Maldacena:1997re,Maldacena:1998re,Witten:1998qj,Aharony:1999ti}.

Considering that asymptotically AdS black hole has a conformal  boundary at infinity, one could deform the boundary metric of a
 black hole and obtain a black hole with deforming horizon, which means  the curvature of the horizon is not a constant value.
 A family of the  hyperbolic AdS black
holes with deforming horizon has recently been constructed analytically \cite{Chen:2015zoa} in four-dimensional  spacetime by using the AdS C-metric \cite{levicivita1917,weyl1917,Plebanski:1976gy}.
In addition,  a class of four-dimensional AdS black holes with noncompact event horizons of finite area is found and called as black bottle, which has a bottle-shaped horizon \cite{Chen:2016rjt}.
Besides analytical method to study the deforming vacuum black hole in four dimensional AdS spacetime,   a family of  deforming solutions with  differential
rotation boundary was constructed  numerically  in \cite{Markeviciute:2017jcp}, including the soliton and black hole. This class of solutions has a
nontrivial boundary metrics that have a dipolar differential rotation profile
\begin{equation}\label{cc3}
 \Omega(\theta) = \varepsilon \cos\theta,
\end{equation}
where  the constant $\varepsilon>0$ is the boundary rotation parameter
and polar angle $\theta$ is  restricted to the interval $(0, \pi)$. It is obvious that
there exists  an anti-symmetric rotation profile with respect to
reflections on the equatorial plane $\theta= \pi/2$. The rotational boundary can yield the  pulling forces, which are maximal at $\theta= \pi/4$ and $3/4\pi$
and could deform the black hole horizon into two hourglass shapes. When the boundary deformation is  larger than  critical parameter $\varepsilon=2$, the norm of Killing vector  $\partial_t$  becomes spacelike for certain regions of $\theta$ when $\varepsilon>2$, which also are called as ergoregions. As a consequence,  both solitons and black holes could develop hair due to superradiance.
In  \cite{Green:2015kur}, the authors found  that spacetimes with ergoregions in AdS may be unstable due to superradiant scattering.
So,  one  could generalize the above conclusions  for that of  nontrivial boundary metrics. Furthermore, a family of deforming vacuum solutions  with  a noncompact, differential rotation boundary metric  was numerically studied in \cite{Crisford:2018qkz}.  With the help of  AdS C-metric, the authors in \cite{Horowitz:2018coe} studied how changes in the boundary metric affect the
shape of the  hyperbolic and
compact AdS black holes. When the matter fields are introduced, one could construct the black holes with deforming horizon in $D=5$ minimal gauged supergravity \cite{Blazquez-Salcedo:2017kig}.

Besides the deforming solutions with dipolar differential rotation boundary,  it will be interesting to see whether there exists the deforming solutions with  multipolar differential rotation boundary.
In the present paper,  we would like to numerically solve Einstein equations and
give  a  family of   deforming black holes with even multipolar differential rotation boundary,
which has the anti-symmetric rotation profile with respect to
reflections on the equatorial plane and  keeps   total angular momentum of black hole to be zero. Especially, considering the configuration of quadrupolar rotation boundary, we obtain the numerical results of  the deforming solitons and black holes.
 Comparing  with the results of dipolar differential rotation, we find  that
 the norm of Killing vector  $\partial_t$  becomes spacelike for certain regions of $\theta$ when $\varepsilon\in(2,2.281)$,
however, black holes with quadrupolar differential rotation  do not develop hair due to superradiance, which  was different from the case of  dipolar rotation.
 Using the isometric embedding of horizon, we can see  the black hole horizon is deformed into  four hourglass shapes.  Furthermore,  we also study the numerical solutions of  entropy and  quasinormal modes, which have the analogous properity   to that
of dipolar rotation boundary in \cite{Markeviciute:2017jcp}.

 The paper is organized as follows. In Sec. \ref{sec2}, we introduce  the model of
  the deforming  black holes with even multipolar differential rotation boundary  and the numerical  DeTurck method.  In Sec. \ref{sec3}, soliton solutions  with quadrupolar differential rotation boundary are constructed numerical, in addition, the numerical results of Kretschman scalar and quasinormal modes is shown. Numerical results of  deforming black holes  with quadrupolar differential rotation boundary are also shown  in Sec. \ref{sec4}.  The conclusion and discussion are given in the
  last section.

%%%%%%%%%%%%%%%%%%%%%%%%%%%%%%%%%%%%%%%%%%%%%%%%%%%%%%%%%%%%%%%%%
%%%%%%%%%%%%%%%%%%%%%%%%%%%%%%%%%%%%%%%%%%%%%%%%%%%%%%%%%%%%%%%%%
\section{Model and numerical method}\label{sec2}
%%%%%%%%%%%%%%%%%%%%%%%%%%%%%%%%%%%%%%%%%%%%%%%%%%%%%%%%%%%%%%%%%
%%%%%%%%%%%%%%%%%%%%%%%%%%%%%%%%%%%%%%%%%%%%%%%%%%%%%%%%%%%%%%%%%
%%%%%%%%%%%%%%%%%%%%%%%%%%%%%%%%%%%%%%%%%%%%%%%%%%%%%%%%%%%%%%%%%
Let us begin with the model of  the four-dimensional Einstein-Hilbert action with a negative cosmological constant $\Lambda$
\begin{align}
 S=\frac{1}{16\pi G}\int \mathrm{d}^4x&\sqrt{-g}\left(R-2\Lambda\right),
 \label{eq:action}
\end{align}
where $G$ is the gravitational constant,
the cosmological constant  is written in terms of the AdS radius $L$ as $\Lambda= -\frac{3}{L^2}$,
$g$ is the determinant of the metric tensor and
 $R$ is the Ricci scalar. The equations of motion derived from (\ref{eq:action}) take  the following form
\begin{align}
\label{eq:EKG1}
G_{ab}\equiv R_{ab}+\dfrac{3}{L^2}g_{ab}=0.
\end{align}
The solution of Einstein equations (\ref{eq:EKG1}),  which can describe  the static spherically symmetric black holes
 with mass, is the well-known AdS-Schwarzschild black hole with the metric given by
\begin{eqnarray}\label{metric11}
 ds^{2} &=& -\left(1-\frac{2M}{r}+\frac{r^2}{L^2}\right)dt^2+\left(1-\frac{2M}{r}+\frac{r^2}{L^2}\right)^{-1}dr^2 +r^2 d\Omega^2,
\end{eqnarray}
where $d\Omega^2$
is the metric on the sphere $S^2$.
Here, the constants  $M$ is the mass of black hole  as measured
from the infinite boundary.
 The horizon radius, denoted by $r_+$, satisfies the
equation
\begin{equation}\label{root}
1-\frac{2M}{r}+\frac{r^2}{L^2}=0,
\end{equation}
and is the largest root, and Hawking
temperature $T_H$  of AdS-Schwarzschild black hole is given by
\begin{eqnarray}
 T_H=\frac{L^2+3 r_+^2}{4\pi L^2 r_+}.
\end{eqnarray}
As near infinity, the metric (\ref{metric11}) is asymptotic to the  anti-de Sitter spacetime,  and boundary
metrics
is conformal
and given by
\begin{equation}\label{boundary}
  ds_\partial^2=r^2(-dt^2+d\theta^2+\sin^2\theta d\phi^2).
\end{equation}
 In order to obtain the new asymptotic Anti-de Sitter solution, the author in \cite{Markeviciute:2017jcp} add differential rotation to the boundary
metric, which is given by
\begin{equation}
   ds_\partial^2=r^2\left(-dt^2+d\theta^2+\sin^2\theta[d\phi+\Omega(\theta)dt]^2\right),
\label{eq:boundary}
 \end{equation}
with  a dipolar differential rotation $\Omega(\theta)=\varepsilon\cos\theta$. Therefore,
the stationary solutions  with boundary
metrics of the form (\ref{eq:boundary}) is the  axisymmetric, and the norm of Killing vector $\partial_t$ is
\begin{equation}
   \|{\partial{t}}\|^2=-1+\frac{\varepsilon^2}{4}\sin^2({2\theta}),
   \label{eq:timelike}
 \end{equation}
with the maximal value at $\theta=\frac{\pi}{4}$.

\begin{figure}[t]
\centering
    \begin{minipage}[t]{0.48\textwidth}
    \includegraphics[width=\textwidth]{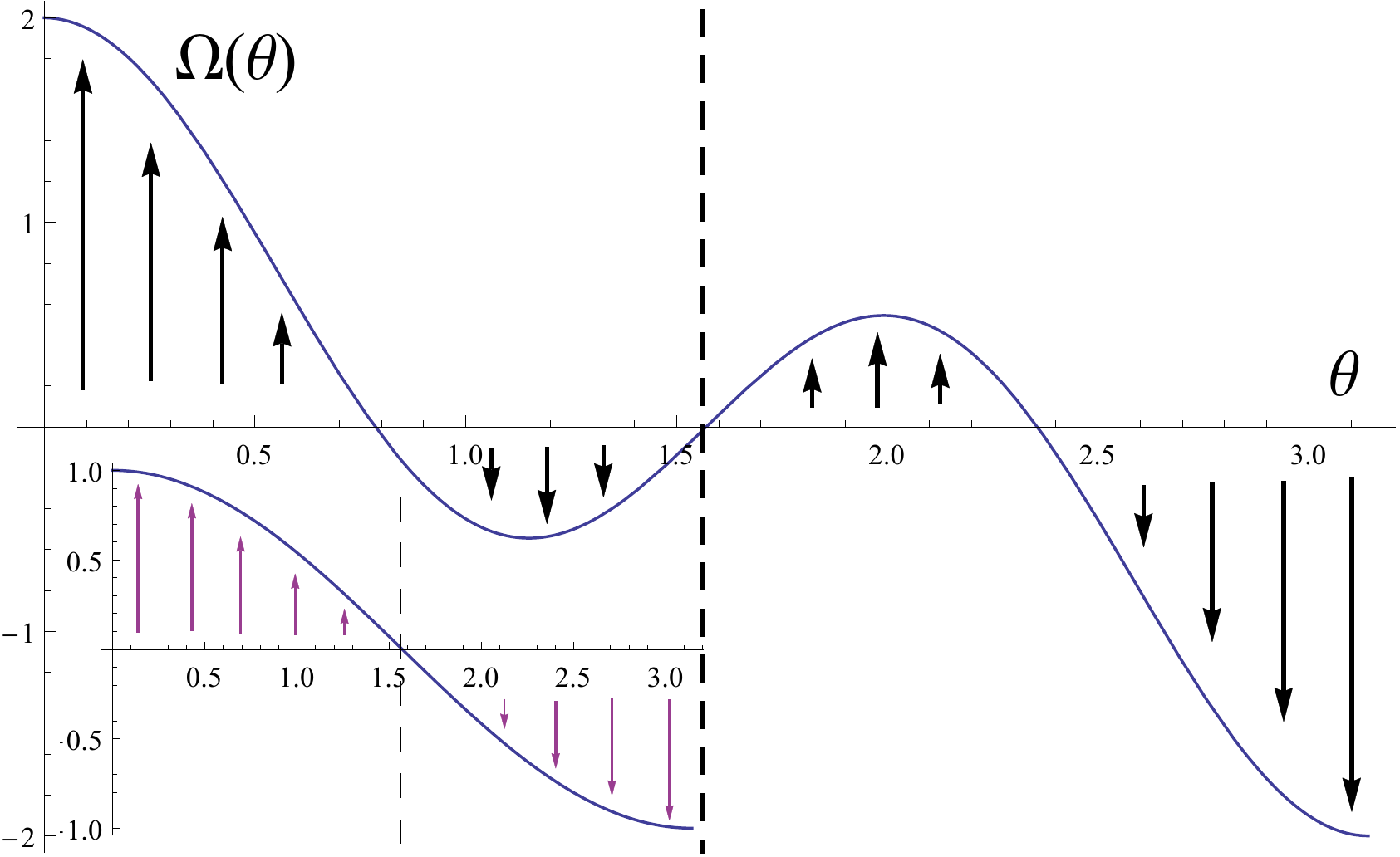}
  \end{minipage}
 \begin{minipage}[t]{.48\textwidth}
   \includegraphics[width=\textwidth]{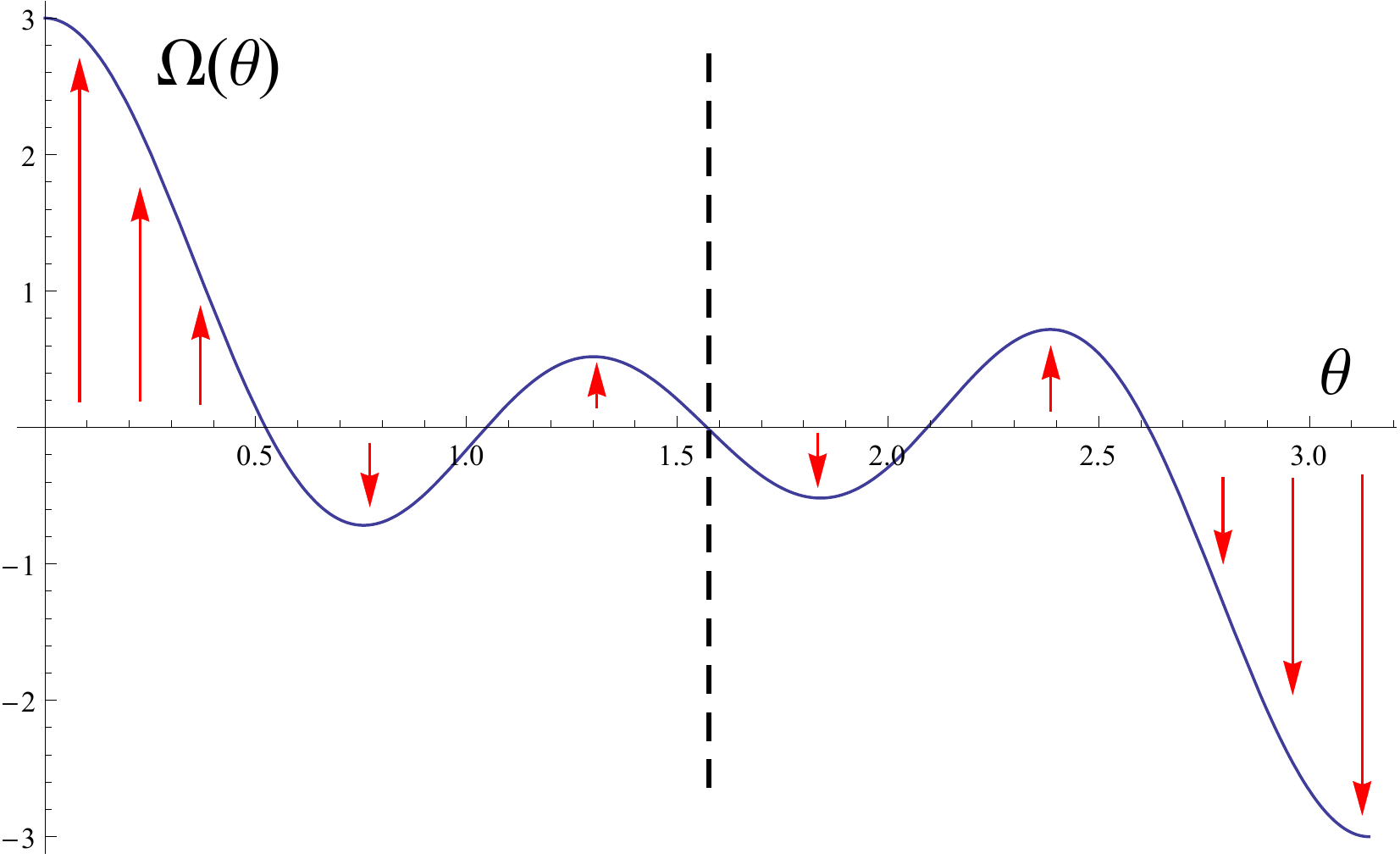}
  \end{minipage}
  \caption{ The differential rotation profile $ \Omega$ as a function of $\theta$ with  $k=4$ (left panel) and $k=6$ (right panel), respectively. In both graphs the arrow lines denote the
orientation of  differential rotation.  The inset in the left panel  shows the  profile of the dipolar differential rotation. }
   \label{fig:pole}
\end{figure}

In order to construct higher even multipolar differential rotation of the conformal boundary,
we also adopt the axisymmetric metric with Kerr-like coordinates \cite{Markeviciute:2017jcp} within the following  ansatz of Killing vector
\begin{equation}
   \|{\partial{t}}\|^2=-1+\frac{\varepsilon^2}{4}\sin^2({k\theta}), \;\;\; k=2,4,6,\cdots,
   \label{eq:timelike}
 \end{equation}
which correspond to the even multipole differential rotations
\begin{equation}\label{dfk}
 \Omega(\theta)=\left\{\begin{array}{c}
\varepsilon\cos\theta,\;\;\;\;\;\;\;\;\;\;\;\;\;\;\;\;\;\;   k=2, \\
\varepsilon(\cos\theta+\cos{3\theta}),  \;\;\;k=4, \\
\frac{\varepsilon}{2}\csc{\theta}\sin{(6 \theta)},\;\;\;\;\;  k=6,
\end{array}
\right.
\end{equation}
where $k=2$ is  the dipolar differential rotation profile,
and the norm of  Killing vector $\partial{t}$ with $ k = 4$ (called the quadrupolar solution) and $ k = 6$ (called the hexapolar solution)
 have  the maximal value at $\theta=\frac{\pi}{8}$ and  $\theta=\frac{\pi}{12}$, respectively.
In  Fig.~\ref{fig:pole}, we draw the graphs of  the differential rotation $ \Omega$ as a function of $\theta$ with  $k=4$ (left panel) and $k=6$ (right panel), respectively. In both graphs the arrow lines denote the
orientation of  differential rotation.  The inset in the left panel of Fig.~\ref{fig:pole}  shows the  profile of the dipolar differential rotation.
We can see that  the even multipoles differential rotation $ \Omega$  are the anti-symmetric functions with respect to
reflections on the equatorial plane $\theta=\pi/2$, which  guarantees that total angular momentum of black hole is zero.

%%%%%%%%%%%%%%%%%%%%%%%%%%%%%%%%%%%%%%%%%%%%%%%%%%%%%%%%%%%%%%%%%%%%
%%%%%%%%%%%%%%%%%%%%%%%%%%%%%%%%%%%%%%%%%%%%%%%%%%%%%%%%%%%%%%%%%%%%
In order to obtain the numerical solution of Einstein  equation (\ref{eq:EKG1}), we use the DeTurck method
\cite{Headrick:2009pv,Wiseman2012,Dias:2015nua}.
By adding a gauge fixing term to Einstein equation, we can obtain a set of  elliptic equations, which are  known as  Einstein-DeTurk equation
\begin{equation}
  R_{ab}+\frac{3}{L^2}g_{ab}-\nabla_{(a}\xi_{b)}=0,
  \label{eq:deturck}
\end{equation}
where $\xi^a=g^{bc}(\Gamma^{a}_{bc}[g]-\Gamma^{a}_{bc}[\tilde{g}])$ is the Levi-Civita connection associated with a reference metric $\tilde{g}$.
It is noted that  reference metric $\tilde{g}$ should be choose to be as same boundary and horizon structure as $g$.

Using   numerical methods for solving these equations of motion,
we could obtain two classes of solutions:
horizonless soliton solutions  with $r_H=0$ and black hole solutions with $r_H >0$.  The soliton solutions can be seen as   deformations
of the global  Anti-de Sitter spacetime, while  black hole solutions closely correspond to  deformations of AdS-Schwarzschild black holes.
For simplify, in our paper we only show the numerical results of the quadrupolar differential rotation  $k=4$, and the cases of higher even-multipole differential rotation have similar behaviour as that of quadrupolar solution.

%%%%%%%%%%%%%%%%%%%%%%%%%%%%%%%%%%%%%%%%%%%%%%%%%%%%%%%%%%%%%%%%%%
\section{\label{subsec:sol}Soliton solutions}\label{sec3}
In this section, it is convenient to compactify both the radial coordinate $r$ and the polar angle coordinate $\theta$
using the change of variables  $r=Ly\sqrt{2-y^2}/(1-y^2)$ and $\sin \theta=1-x^2$,
which implies that the new  radial coordinate $y  \in [0,1]$ and polar angle coordinate $x  \in [0,1]$. Thus the inner and outer boundaries of the shell
are fixed at $y = 0$ and $y = 1$, respectively.
In order to solve  the above coupled equations   (\ref{eq:EKG1}) numerically with  a quadrupole  differential rotation (\ref{dfk}),  we choose the  ansatz of solitonic solutions as
\begin{multline}
\mathrm{d}s^2=\frac{L^2}{(1-y^2)^2}\Bigg\{-U_1\,\mathrm{d}t^2+\frac{4\, U_2\,\mathrm{d}y^2}{2-y^2}+y^2 (2 - y^2) \Bigg[\frac{4\,U_3}{2-x^2}\left(\mathrm{d}x+\frac{x}{y} \sqrt{2-x^2}\,\,U_4\, \mathrm{d}y\right)^2 \\+(1-x^2)^2 U_5\,\left(\mathrm{d}\phi+y x\sqrt{2-x^2}\left(-2+4x^2(2-x^2)\right)\,U_6\,\mathrm{d}t\right)^2 \Bigg]\Bigg\}\,,
\label{eq:ansatzsol}	
\end{multline}
where the functions $U_i ~ ( i= 1,2,3,4,5,6)$  depend on the variables  $x$  and  $y$.  When $U_1 = U_2 = U_3 = U_5 = 1$ and
$U_4 = U_6 = 0$, the metric (\ref{eq:ansatzsol}) can reduce to  Anti-de Sitter spacetime in global coordinates.

Before  numerically solving the differential equations instead of seeking the analytical solutions, we should obtain the asymptotic behaviors of the  six functions $U_i ~ ( i= 1,2,3,4,5,6)$, which are equivalent to know the boundary conditions we need.
%Considering  the  properties of
%Kerr black holes with excited state scalar hair, we will still use  the  %boundary conditions by following the same steps as  the ground state   given in %Refs. \cite{Herdeiro:2014goa, Herdeiro:2015gia}.
Because  the solutions have properties of polar angle reflection symmetry $\theta\rightarrow\pi-\theta$ on the equatorial plane,   it is convenient to consider the
 coordinate range $\theta \in [0,\pi/2] $, i.e. $x \in [0,1] $. So,  we require  the functions  to satisfy the following Neumann boundary conditions
on the equatorial plane $x=0$
\begin{equation}
\partial_x U_i(0,y)=0,  \;\;\;i=1,2,3,4,5,6,
\end{equation}
and  set  axis boundary conditions at $x=1$,  where  regularity must be imposed    Dirichlet boundary conditions on
$U_4$
\begin{equation}\label{abc}
U_4(1,y)=0,
\end{equation}
and Neumann boundary conditions on the other functions
\begin{equation}\label{abc}
\partial_x U_1(1,y)=\partial_x U_2(1,y)=\partial_x U_3(1,y)=\partial_x U_5(1,y)=\partial_x U_6(1,y)=0.
\end{equation}
Moreover, expanding the equations of motion near  $x=1$  gives the condition $U_3(1,y)=U_5(1,y)$.
In addition,
the asymptotic behaviors near the conformal boundary $y=1$ are
\begin{eqnarray}
  U_4(x,1) =  0 ,\;\;\;  U_6(x,1)  = \varepsilon, \nonumber\\
 U_1(x,1)= U_2(x,1) =U_3(x,1) =U_5(x,1) =1,
\end{eqnarray}
and finally,
by expanding the equations of motion near
$y = 0$ as a power series in $y$, we have
 \begin{equation}
\partial_y U_i(x, 0)   = 0,  \;\;\;i=1,2,3,4,5,6.
\end{equation}
Note that in the center $y=0$ of   soliton solutions,   not all of  the values of  $U_i(x, 0)  $ are the constants   independent of     polar angle $ x$. According to asymptotic behaviors near $y=0$, we obtain that
\begin{equation}\label{center}
 U_1(x,0)=c_1,\;\;\; U_5(x,0)=c_5,\;\;\; U_6(x,0)=c_6,\;\;\;
\end{equation}
\begin{equation}\label{center}
 U_2(x,0)=U_2^{(0)}(x) ,\;\;\; U_3(x,0)=U_3^{(0)}(x),\;\;\; U_4(x,0)=U_4^{(0)}(x) ,\;\;\;
\end{equation}
where the parameters $c_1$, $c_5$ and $c_6$ can take
arbitrary constant value, and  $U^{(0)}_2$, $U^{(0)}_3$ and $U^{(0)}_4$ are the functions dependent of $x$.

With the above boundary conditions, the ansatz of   metric has the boundary forms of Eq. (\ref{eq:boundary}) with  the quadrupolar differential rotation  $\Omega(\theta)= \varepsilon(\cos\theta+\cos{3\theta})$.
We can choose the reference metric $\tilde{g}$  given by the line element (\ref{eq:ansatzsol}) with
$U_4=0, U_6=\varepsilon$ and $U_1=U_2=U_3=U_5=1$.

\begin{figure}[t]
\centering
  \begin{minipage}[t]{0.49\textwidth}
    \includegraphics[width=\textwidth]{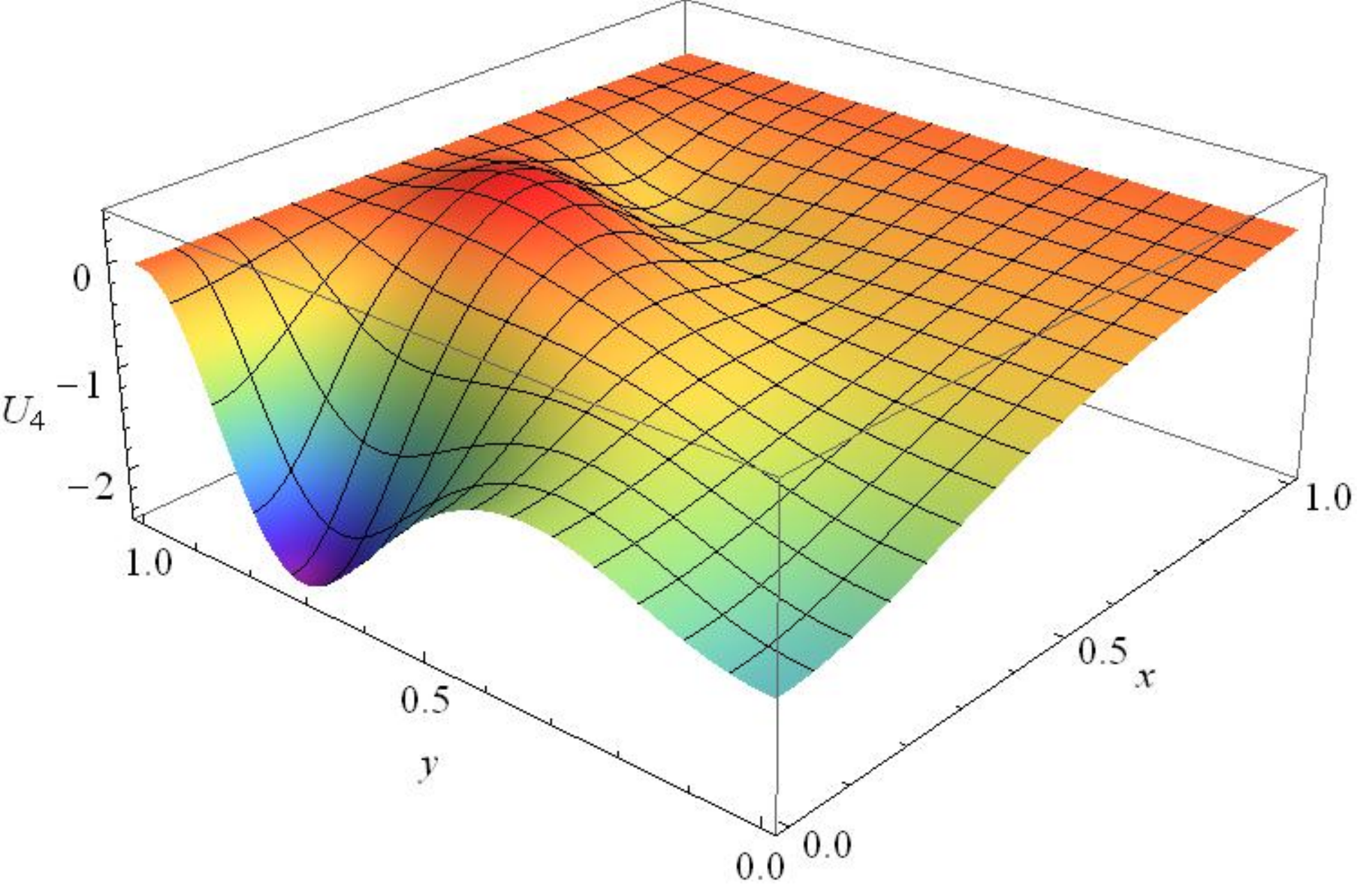}
  \end{minipage}
    \begin{minipage}[t]{0.49\textwidth}
    \includegraphics[width=\textwidth]{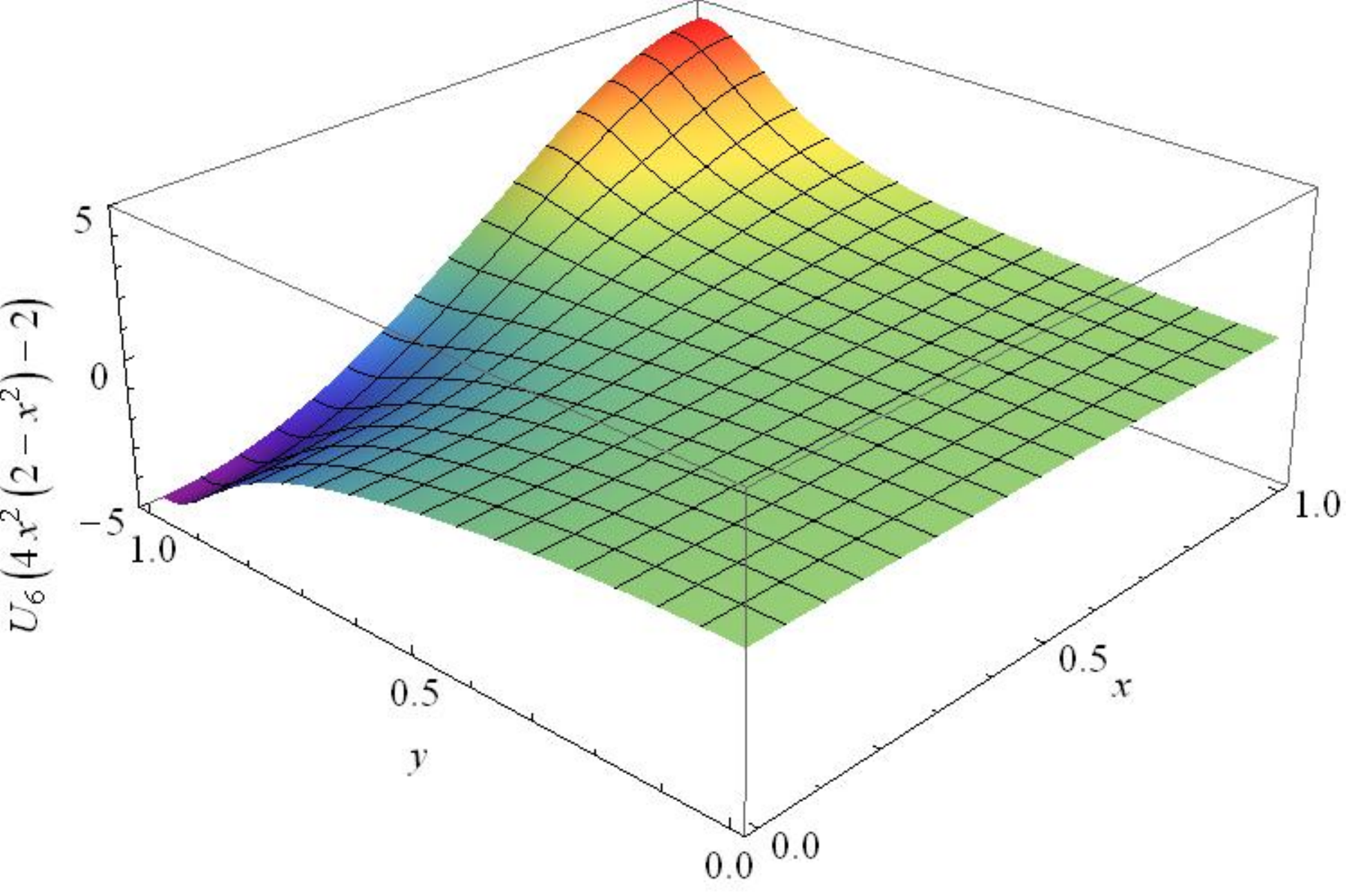}
  \end{minipage}
    \begin{minipage}[t]{0.49\textwidth}
    \includegraphics[width=\textwidth]{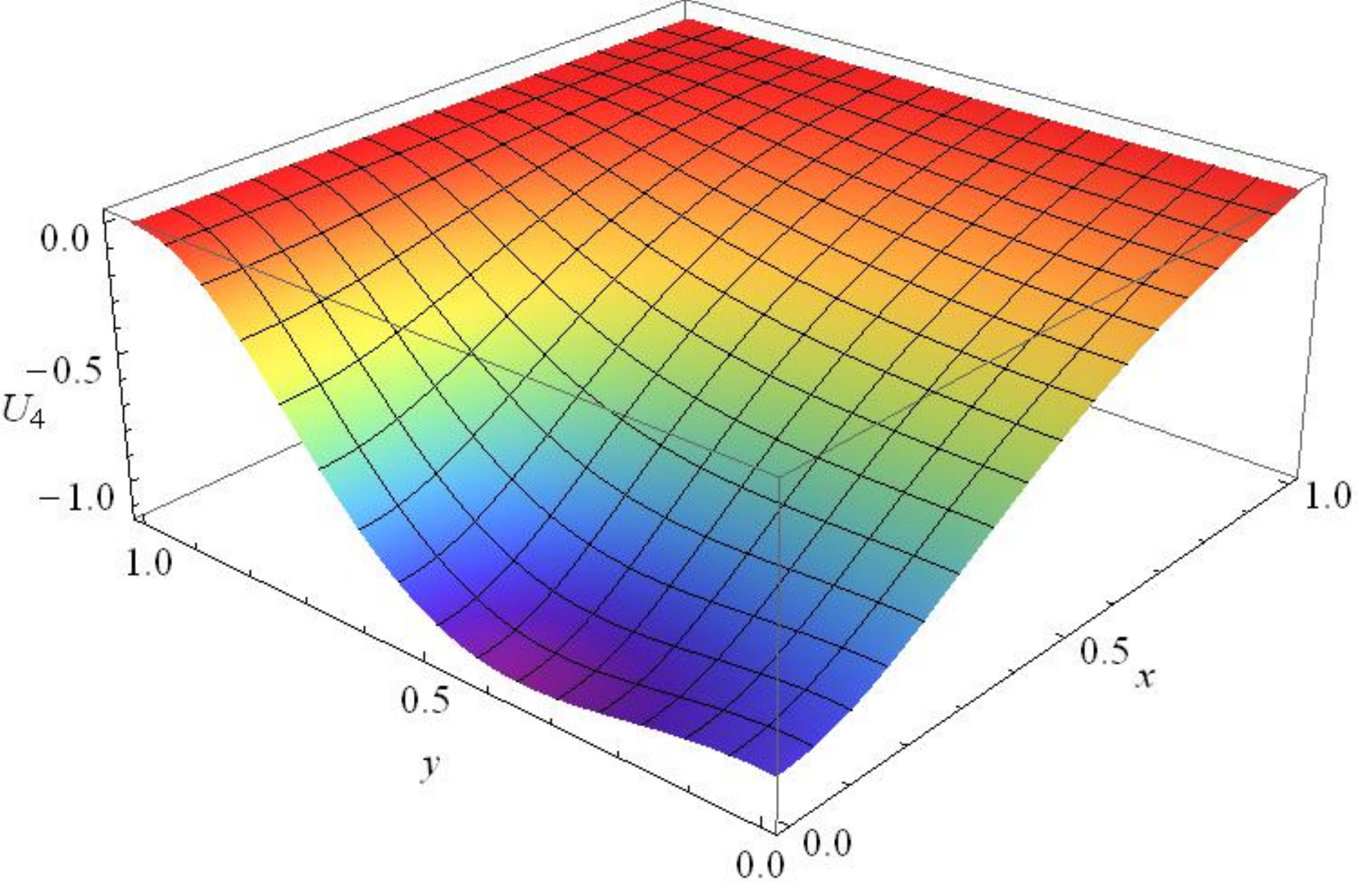}
  \end{minipage}
    \begin{minipage}[t]{0.49\textwidth}
    \includegraphics[width=\textwidth]{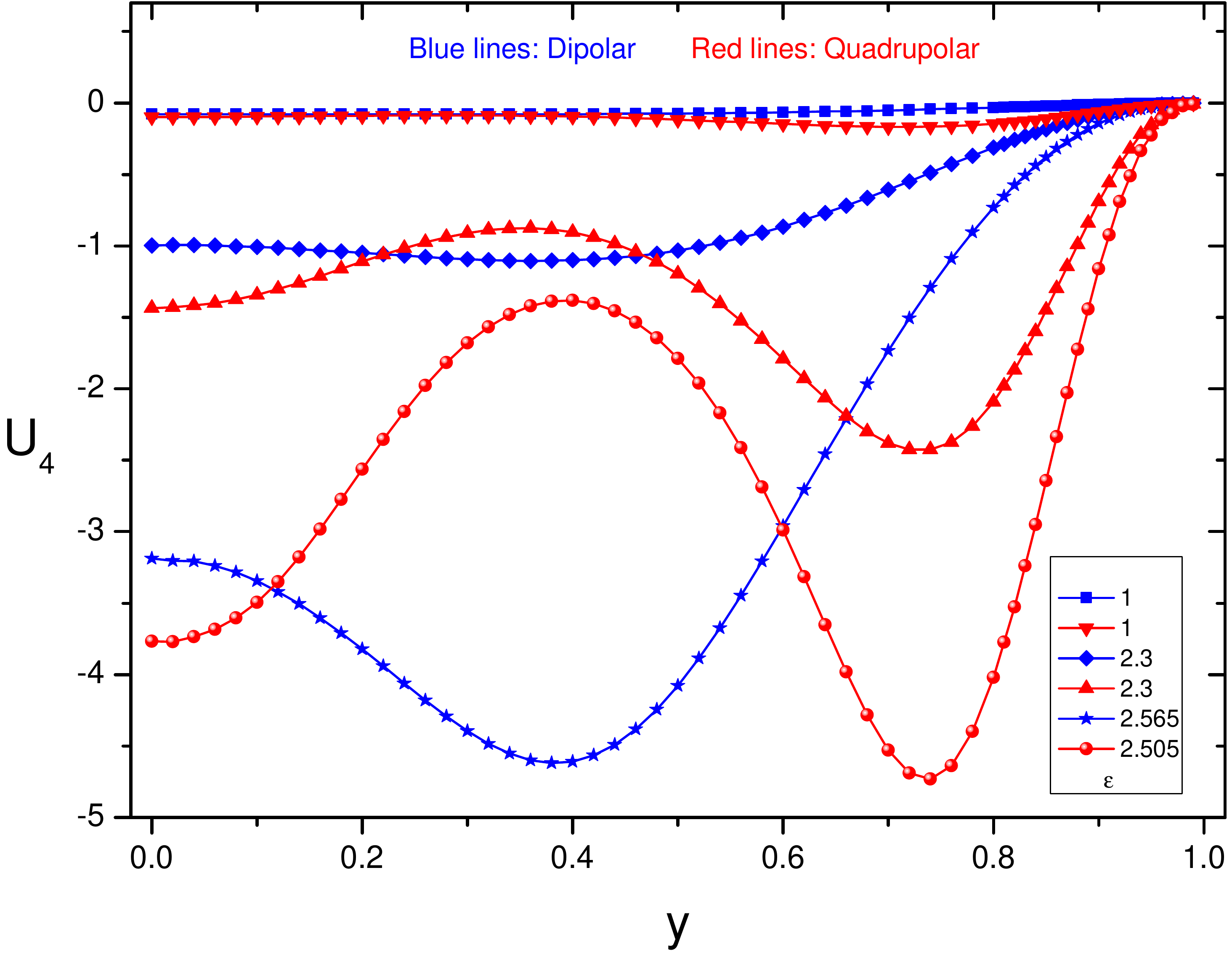}
  \end{minipage}
  \caption{On the top left we show $U_4$ and on the top right $U_6 (4x^2(2-x^2)-2)$  of soliton solution with the  quadrupolar boundary rotation, and the two figures on the top have $\varepsilon=2.3$. On the bottom left we show $U_4$  of soliton solution with the  dipolar boundary rotation for $\varepsilon=2.3$ and on the bottom right the distribution of $U_4$ as a function of  the $y$ coordinate  in  the equatorial plane  at $x= 0$ for various
 rotation parameters $\varepsilon$.}
  \label{fig:sol}
\end{figure}

In the top of Fig.~\ref{fig:sol}, we show  the typical soliton  result of our numerical code for $U_4$  in the  left panel and  $U_6 (4x^2(2-x^2)-2)$ in the  right panel with the  quadrupolar boundary rotation, and the two figures in the top have the same parameter $\varepsilon=2.3$.  In order to explore the influence of the different boundary rotations on the metric,   in the bottom left we show $U_4$   with the  dipolar boundary rotation for the same paramter $\varepsilon=2.3$.  Furthermore, in the bottom right the distributions of $U_4$ as a function of  the $y$ coordinate  at  the equatorial plane    $x= 0$ for various
 rotation parameters $\varepsilon$ are shown, in this plot the curves $U_4$ with $k=2$ and $k=4$ are denoted by blue and red lines,  respectively.
Comparing with the results of the dipolar boundary rotation, we can see that the curve of $U_4$  with the  quadrupolar boundary rotation have more twists and turns than that with the the  dipolar boundary rotation,   and  the minimum value  of $U_4$  with the  quadrupolar boundary rotation is larger than that with the  dipolar boundary rotation.

According to the numerical results, we find there exists stationary axisymmetric soliton solutions  for    $\varepsilon < \varepsilon_c=2.518$, where $\varepsilon_{c}$ is the maximal value  and smaller than the value of dipolar boundary rotation.
Moreover,  the soliton solutions for each value of $\varepsilon\in(2.281,2.518)$ have two branchs.

%%%%%%%%%%%%%%%%%%%%%%%%%%%%%%%%%%%%%%%%%%%%%%%%%%%%%%%%%%%%%%%%%%%%
\subsection{\label{subsubsec:kresc}Kretschman scalar}
\begin{figure}[t]
\centering
  \begin{minipage}[t]{.495\textwidth}
    \includegraphics[width=\textwidth]{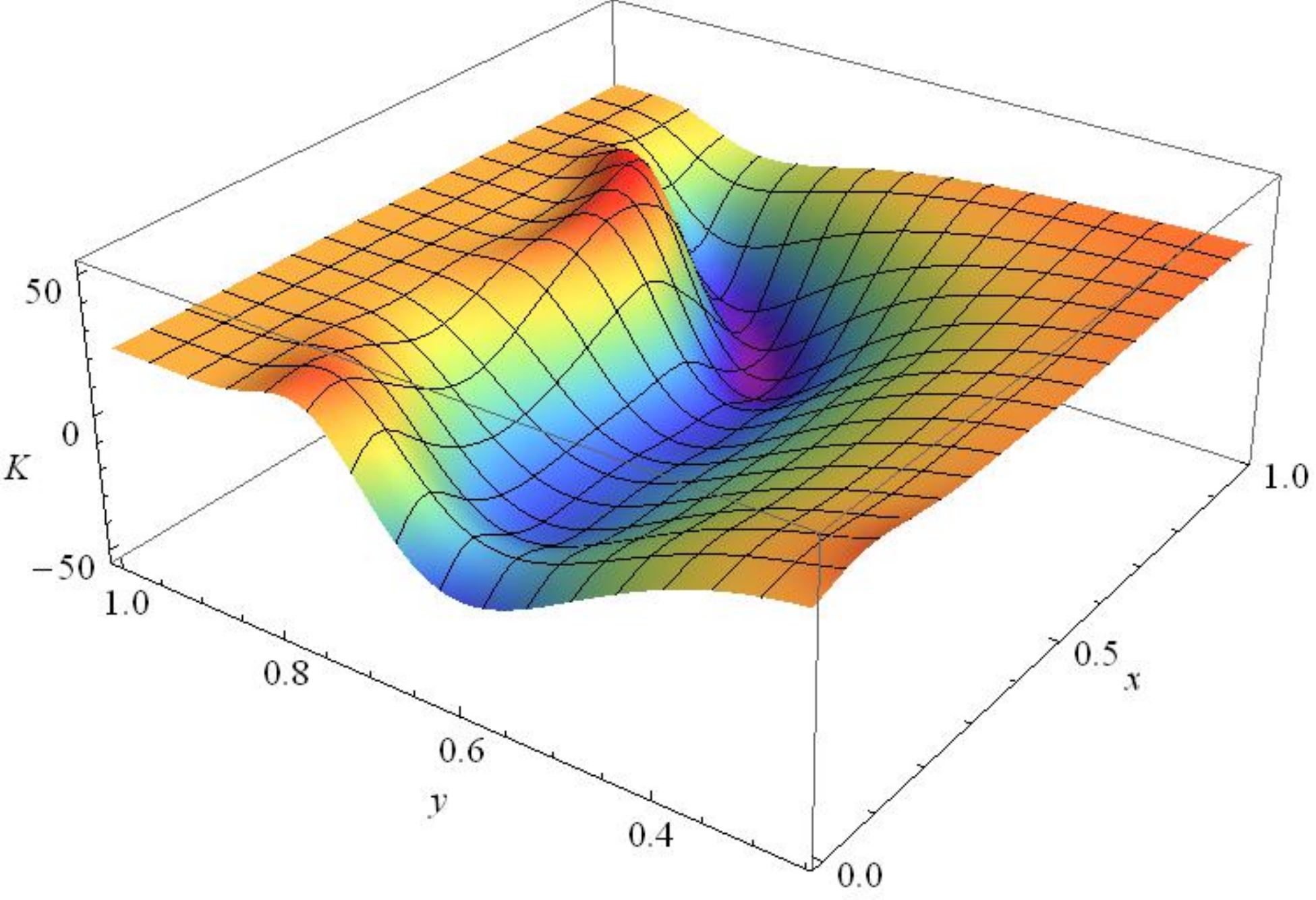}
  \end{minipage}
    \begin{minipage}[t]{.495\textwidth}
    \includegraphics[width=\textwidth]{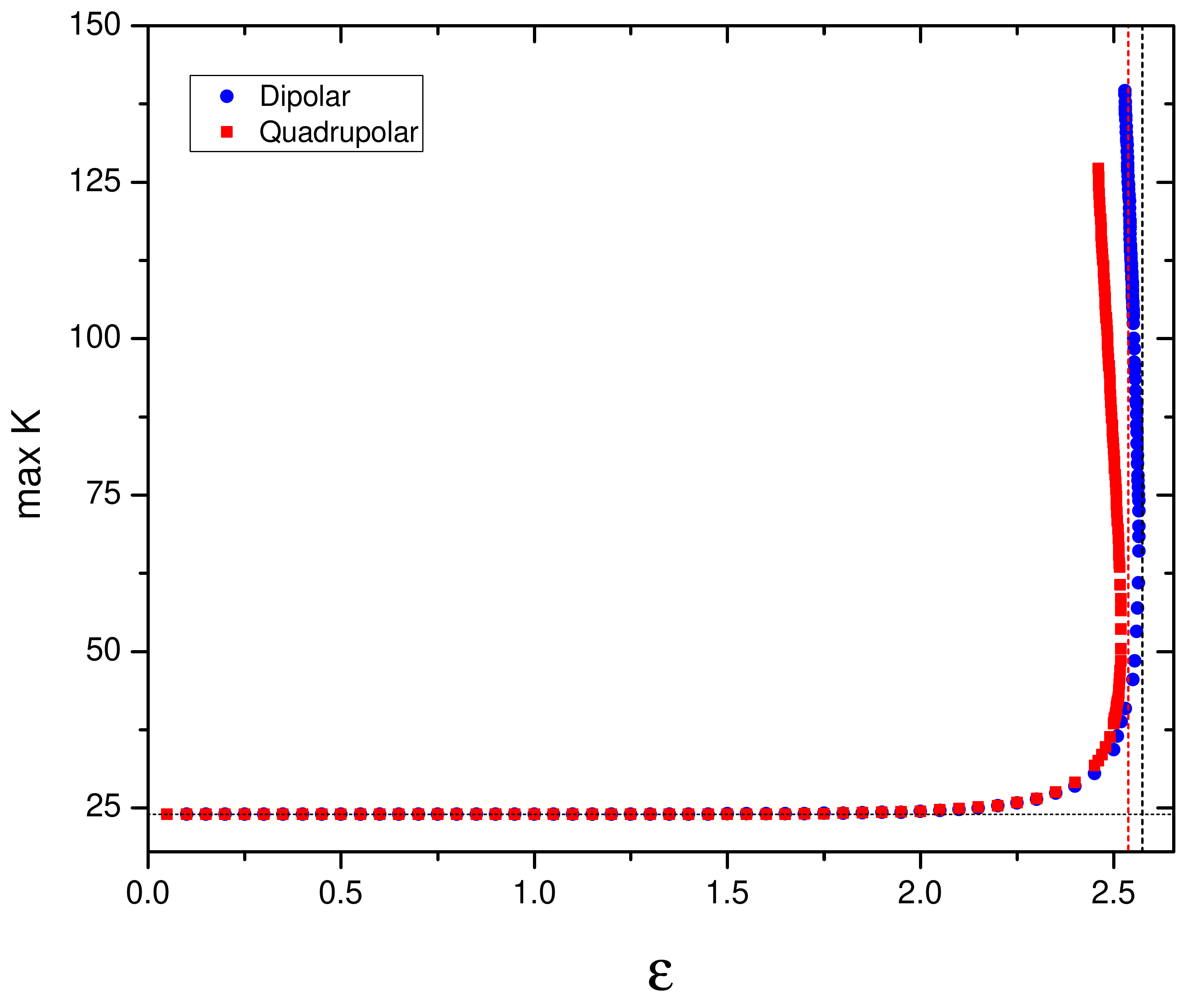}
  \end{minipage}
  \caption{\textit{Right}: The distribution of  Kretschmann scalar  as a function of  $x$ and $y$ coordinate  with the rotation parameter $\varepsilon=2.518$.                                                                                                                                                                                               \textit{Left}: The maximum of the Kretschmann scalar as a function of the rotation parameter for the soliton solutions with  $k=2,~4$, represented by the   blue  and  red lines, respectively.  The  vertical   black and red  dashed gridlines indicate the  $\varepsilon=2.565$ and $\varepsilon=2.518$ maximum value, respectively,
and  the horizon  dashed black line is  the value of $K=24/L^4$ for AdS$_4$ spacetime.}
  \label{fig:Kretschmann}
\end{figure}
When one obtains a solution of Einstein equation, it is very important to know whether the spacetime of solution is
regular or not. Ricci scalar is the simplest curvature invariant of a Riemannian manifold. But, considering that  Ricci tensor $R$ in our model is
$R=2\Lambda$, we need to choose the another invariant  which can indicate the flatness of a chosen manifold. In general, one of the most useful ways is to check for the
finiteness of the Kretschmann scalar,  which sometimes is also called Riemann tensor
squared and written as
\begin{equation}
  K=R_{\alpha\beta\gamma\delta}R^{\alpha\beta\gamma\delta},
\label{eq:KRECUR}
\end{equation}
where $R_{\alpha\beta\gamma\delta}$ is the Riemann curvature tensor. Because it is a sum of squares of tensor components, Kretschmann scalar is a quadratic invariant.

%%%%%%%%%%%%%%%%%%%%%%%%%%%%%%%%%%%%%%%%%%%%%%%%%%%%%%%%%%%%%%%%%%%%%%
~~~Numerical results are presented in Fig~\ref{fig:Kretschmann}. In the left panel we present the Kretschmann scalar $K$ for the quadrupolar solution  with the boundary parameter $\varepsilon\simeq\varepsilon_{c}$, and it is obvious that the spacetime is not flat.
In the right panel,  we exhibit   the maximum of the Kretschmann scalar $K$
versus the rotation
parameter for the soliton solutions  with  $k=2,~4$, represented by the   blue  and  red lines, respectively, and the critical values of the rotation
parameter $k=2$ and $k=4$ are indicated by the vertical black  and red dashed gridlines, respectively. From the figure, there exits the growth of Kretschmann scalar  in  the large branch of quadrupolar rotation solutions, which indicates the formation of a curvature singularity similar to the case of $k=2$ dipolar rotation.

\subsection{\label{subsubsec:QNM}Quasinormal modes}
%%%%%%%%%%%%%%%%%%%%%%%%%%%%%%%%%%%%%%%%%%%%%%%%%%%%%%%%%%%%%%%%%%%%
\begin{figure}[t]
\centering
    \begin{minipage}[t]{0.6\textwidth}
    \includegraphics[width=\textwidth]{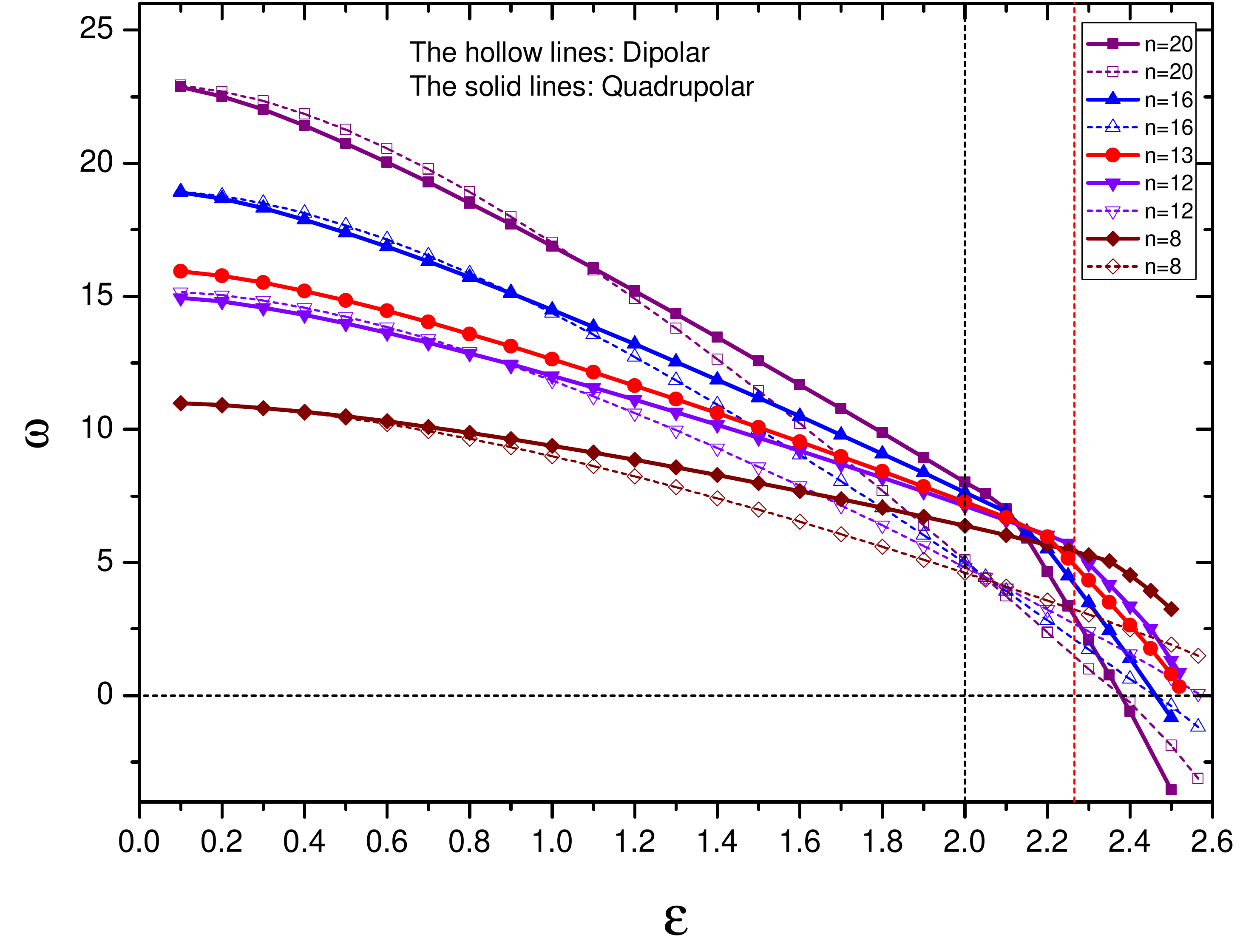}
  \end{minipage}
  \caption{The normal mode frequencies $\omega$  as a function of the rotation parameter $\varepsilon$.
 The solid lines and dashed hollow lines   stand for the quadrupolar  and dipolar  solutions, respectively. The  vertical   black and red  dashed gridlines indicate the values of  $\varepsilon=2$ and $\varepsilon=2.281$, respectively.}
   \label{fig:SoStability}
\end{figure}
In this subsection to study the linear stability of soliton solutions  with the quadrupolar rotation
  $k=4$, we will investigate the quasinormal modes (QNMs),  which are characteristic to the background spacetimes.  Following the method  in papers~\cite{Dias:2013sdc,Cardoso:2013pza,Markeviciute:2017jcp},
we consider a  free, massless scalar field, obeying a massless Klein-Gordon
equation
\begin{equation}
\nabla^2\chi=\frac{1}{\sqrt{-g}}\partial_\mu(\sqrt{-g}g^{\mu\nu}\partial_\nu \chi) =0,
\label{eq:box}
\end{equation}
where scalar
field could be separated into the standard form
\begin{equation}\label{qiuxie1}
\chi(t,x,y,\phi)=\tilde{\chi}(x,y)e^{-i\,\omega\,t+ i n \phi}, \,\;\;\; n=\pm1,\pm2,  \cdots ,
\end{equation}
where the constant $\omega$ is the frequency of the complex scalar field and $n$ is the azimuthal harmonic index.
With the  ansatze of the soliton metric ~(\ref{eq:ansatzsol}), the scalar
field could be further
decomposed into
\begin{equation}
\tilde{\chi}(x,y)=y^{n} (1-y^2)^3 (1-x^2)^{n}\psi(x,y),
\end{equation}
where the powers of $x$ and $y$ were chosen to make function $\psi(x,y)$ to be regular at the origin. In addition, the boundary condition at $y=1$
is given by
\begin{equation}
\partial_y\psi(x,y)=-|n|\, \psi(x,y).
\end{equation}
At $y=0$ and  $x=\pm1$,  we require that the function $\psi(x,y)$ approaches the homogeneous
solution with Neumann boundary conditions.

In the Fig~\ref{fig:SoStability},
we plot  the normal mode frequencies $\omega$  as a function of the rotation parameter $\varepsilon$   for the corresponding  values of  $n$, represented by solid lines. Meanwhile,  we also show the numerical results of QNMs studied in \cite{Markeviciute:2017jcp}, represented by dashed lines. We can see that  normal mode frequencies  $\omega$ with $n\leq13$ are always positive modes in the spectrum of perturbations, while, the frequency becomes negative at a specific value of $\varepsilon$ when $n\geq n_c=14$. One can expect some branches of soliton solution  with scalar hair $\chi$ condensation can be found.
Comparing  with the results of the  dipolar  differential
rotation $k=2$, we see that  the azimuthal harmonic index $n_c=14$  of the  quadrupolar  differential
rotation is larger than the index $n_c=13$  of the  dipolar  differential
rotation.

%%%%%%%%%%%%%%%%%%%%%%%%%%%%%%%%%%%%%%%%%%%%%%%%%%%%%%%%%%%%%%%%%%%%
\section{\label{sec:bla}Black hole solutions}\label{sec4}
\begin{figure}[t]{\label{bhview}}
\centering
  \begin{minipage}[t]{0.49\textwidth}
    \includegraphics[width=\textwidth]{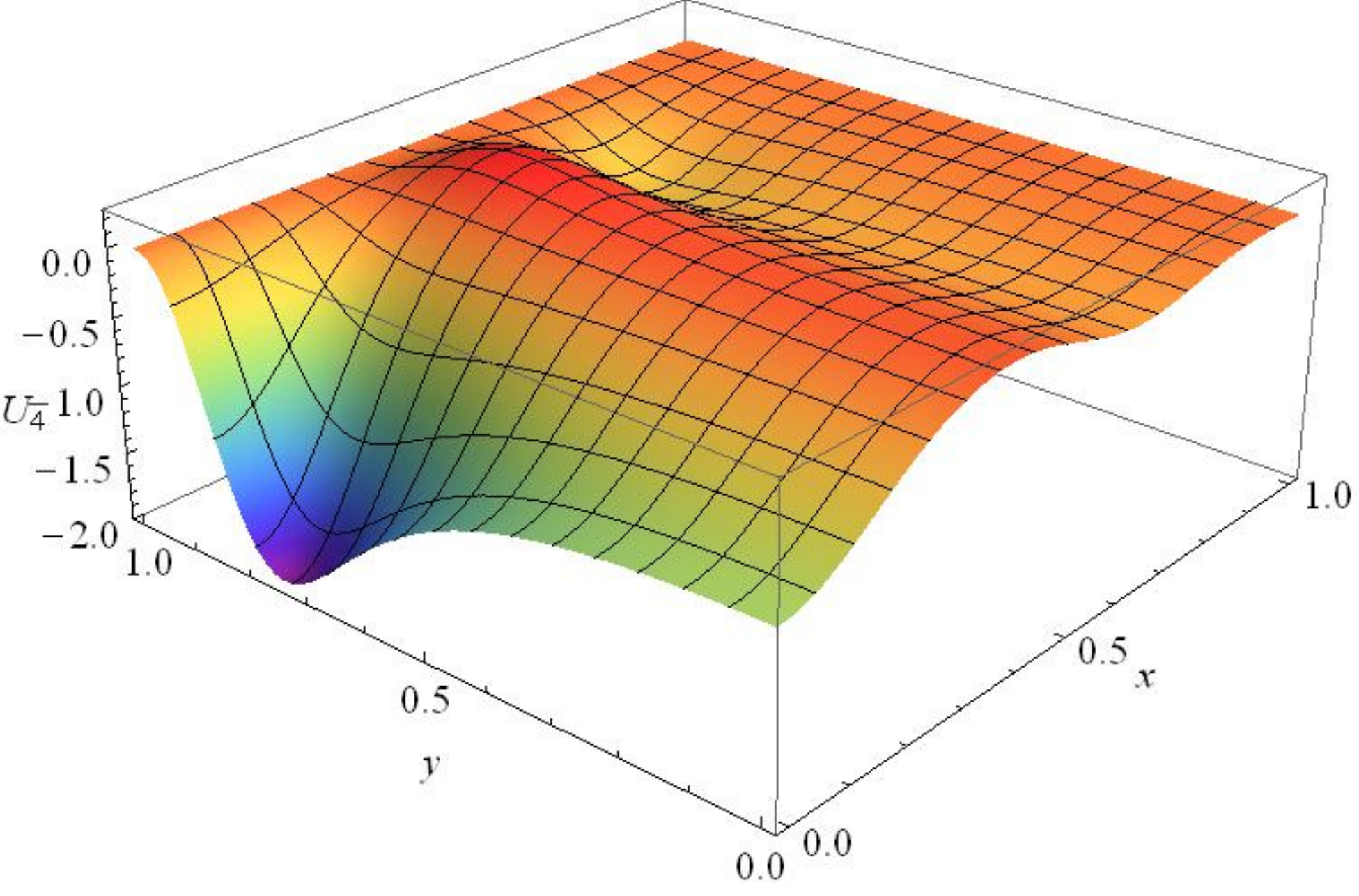}
  \end{minipage}
  \hfill
    \begin{minipage}[t]{0.5\textwidth}
    \includegraphics[width=\textwidth]{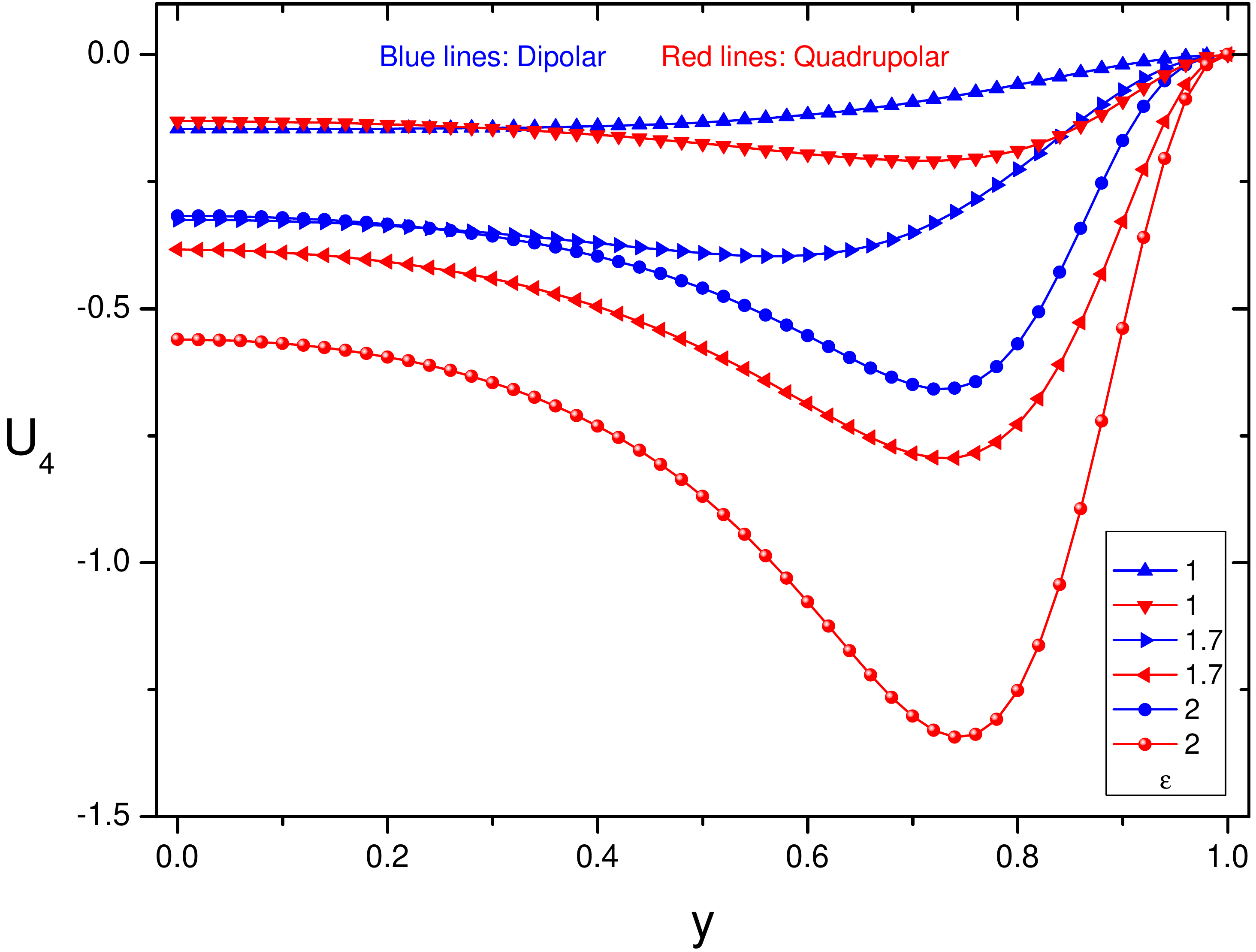}
  \end{minipage}
    \begin{minipage}[t]{0.49\textwidth}
    \includegraphics[width=\textwidth]{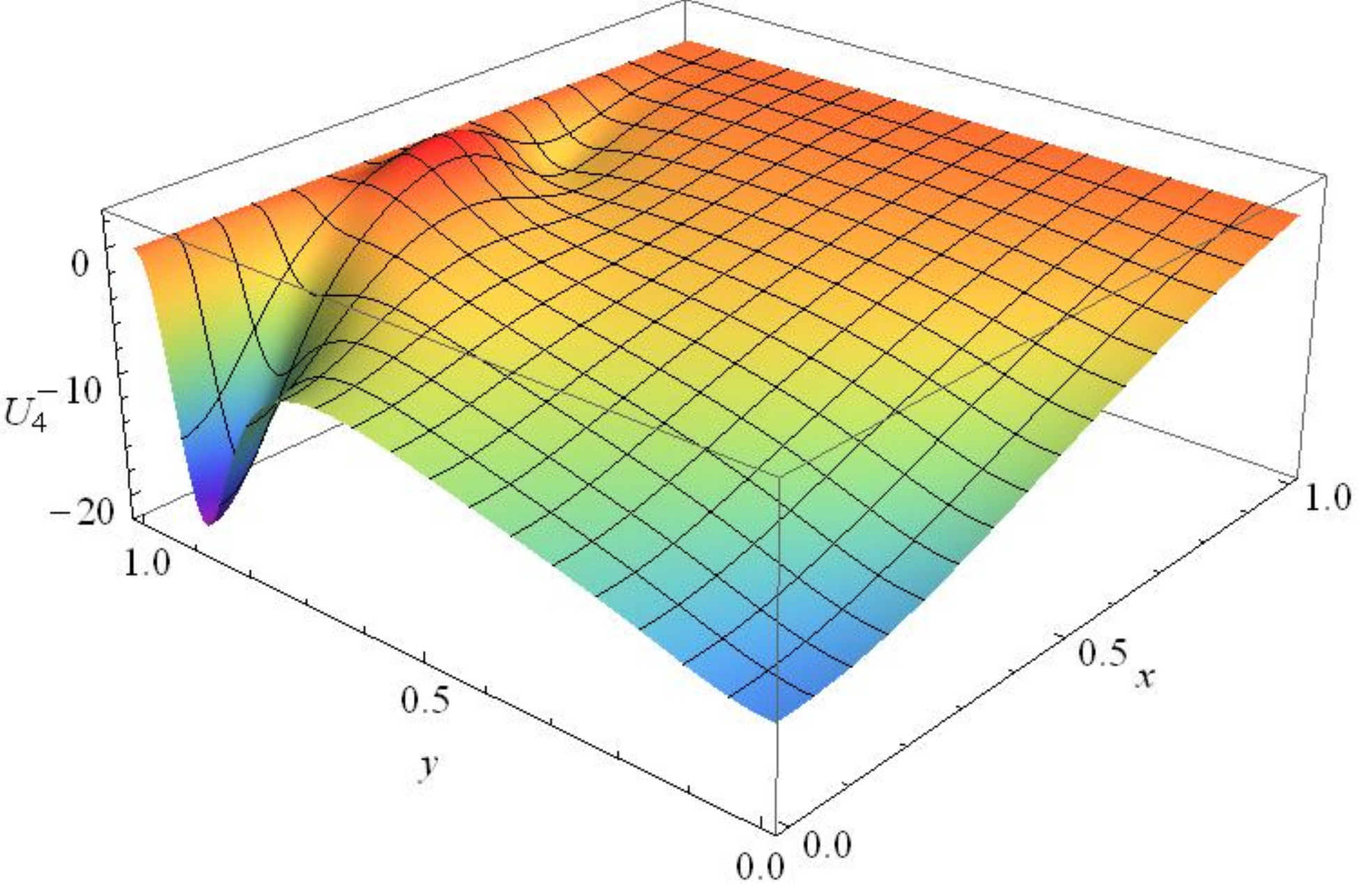}
  \end{minipage}
  \hfill
    \begin{minipage}[t]{0.5\textwidth}
    \includegraphics[width=\textwidth]{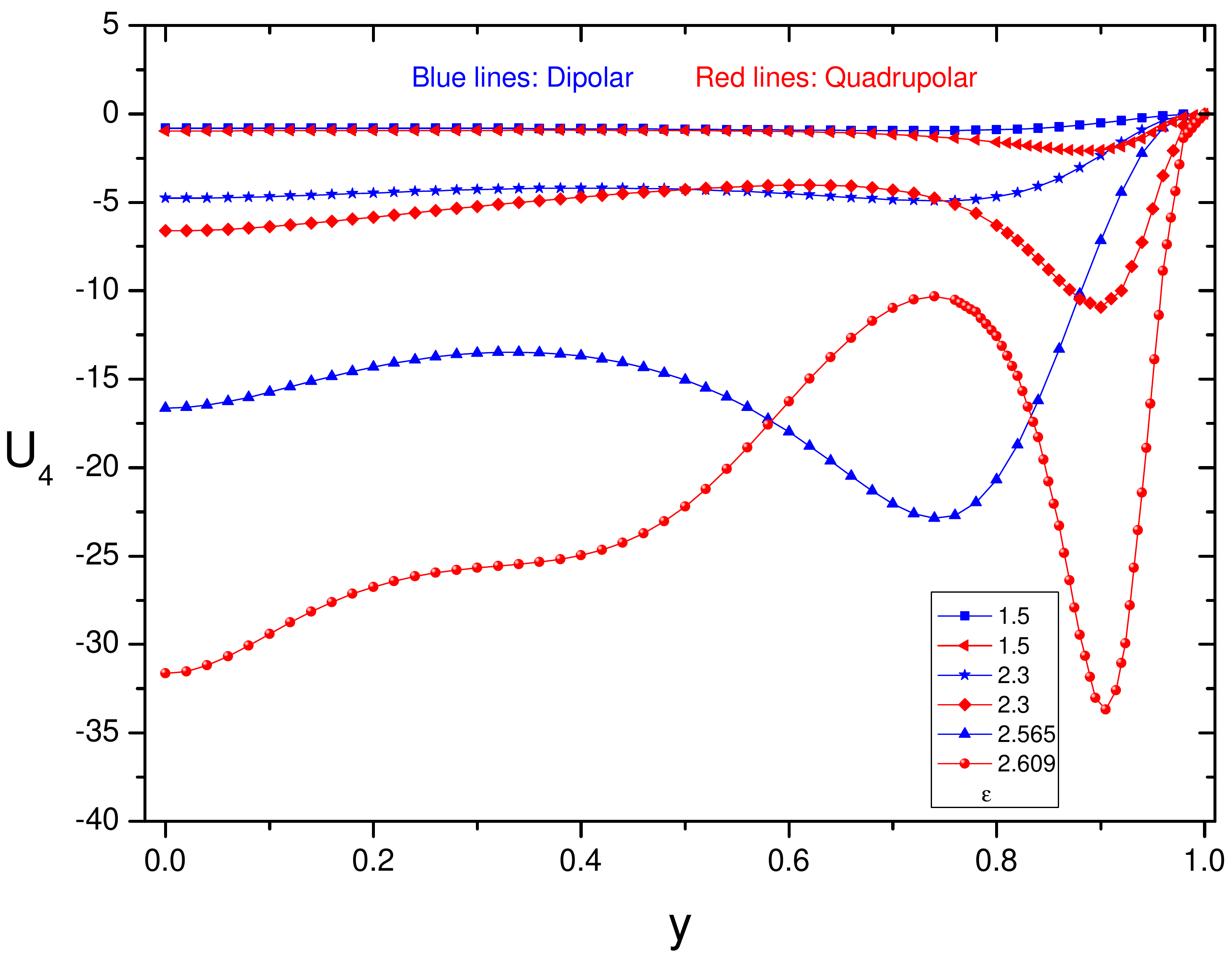}
  \end{minipage}
  \caption{\textit{Top left}: The distributions of $U_4$ as a function of  $x$ and $y$  for large black hole with  quadrupolar rotation parameter $\varepsilon=2.215$. \textit{Top right}: For large black hole, the distributions of $U_4$ as a function of   $y$   at  the equatorial plane for various
values of $\varepsilon$. \textit{Bottom left:} The distributions of $U_4$ for small black hole with  quadrupolar rotation parameter $\varepsilon=2.5$. \textit{Bottom right:} For small black hole, the distributions of $U_4$ as a function of   $y$   at  the equatorial plane for various
values of $\varepsilon$. In  both  right panels,  the red and blue lines correspond to quadrupolar  and dipolar  rotation, respectively.}
  \label{fig:BH}
\end{figure}
In this section, it is convenient to compactify  the radial coordinate $r$ and  polar angle coordinate $\theta$
with the change of variables   $r=L y_p/(1-y^2)$  and $\sin \theta=1-x^2$, respectively.
In order to obtain the black hole solutions with quadrupolar  differential
rotation, we consider the  ansatz of metric
\begin{subequations}
\begin{multline}
\mathrm{d}s^2=\frac{L^2}{(1-y^2)^2}\Bigg\{-y^2\Gamma^+(y)U_1\mathrm{d}t^2+\frac{4\,y_p^2 U_2\,\mathrm{d}y^2}{\Gamma^+(y)}\\+y_p^2 \Bigg[\frac{4\,U_3}{2-x^2}\left(\mathrm{d}x+yx \sqrt{2-x^2}\left(-2+4x^2(2-x^2)\right)\,U_4\, \mathrm{d}y\right)^2\\+(1-x^2)^2 U_5\,\left(\mathrm{d}\phi+y^2x \sqrt{2-x^2}\left(-2+4x^2(2-x^2)\right)U_6\,\mathrm{d}t\right)^2 \Bigg]\Bigg\},
\label{eq:ansatzbh}
\end{multline}
\noindent with
\begin{equation}
\Gamma(y)=(1- y^2)^2 + y_p^2 (3 - 3 y^2 + y^4)\,,\quad\text{and}\quad \Gamma^+(y)= \Gamma(y) \delta + y_p^2 (1 -\delta)\,,
\end{equation}
\end{subequations}
where the functions $U_i(x,y)~( i= 1,2,3,4,5,6)$  depend on the variables  $x$  and  $y$.
Providing that $U_4=U_6=0 $ and $U_1=U_2=U_3=U_5=\delta=1$, the metric~(\ref{eq:ansatzbh}) reduces to the Schwarzschild-AdS black hole.

In the ansatz~(\ref{eq:ansatzbh}), we have three parameters: $\varepsilon$, which sets the amplitude of the boundary rotation, and $(y_p,\delta)$, which determine the black hole temperature. The Hawking temperature of the  black hole with quadrupolar differential rotation is given by
\begin{equation}\label{metric}
  T=\frac{1}{4\pi}\sqrt{-g^{tt}g^{\alpha\beta}\partial _{\alpha}g_{tt}\partial_{\beta}g_{tt}}\mid_{y=0}=\frac{(\delta+2{\delta}y_p^2+y_p^2)}{4{\pi}y_p},
\end{equation}
where the parameter $\delta$ is introduced to control the temperature  to any values.
If $\delta =1$, the temperature has a minimum at $y_p=1/\sqrt{3}$, coinciding with the minimal temperature of a Schwarzschild-AdS$_4$, occurring at $T_{c}\equiv\sqrt{3}/(2\pi)\approx 0.2757$.
It is obvious that the black hole temperature  with quadrupolar differential rotation in (\ref{metric}) is same as that with dipolar differential rotation.
When  $\delta$ have  a fixed value,  one can obtain two solutions with the same temperature, which we call as large black holes with larger $y_p$, and small black holes with smaller $y_p$.

The boundary condition is similar to the soliton case. At $x=0$ and $y=0$,  the functions $U_i$ satisfy the following Neumann boundary conditions
\begin{equation}
\partial_x U_i(0,y)=\partial_y U_i(x,0)=0,  \;\;\;i=1,2,3,4,5,6,
\end{equation}
and at axis boundary  $x=1$ , we require that regularity must be imposed with $U_4=0, U_3=U_5$, and $\partial_x U_1=\partial_x U_2=\partial_x U_3=\partial_x U_5=\partial_x U_6=0$.
At the conformal boundary $y=1$, we set $U_4=0,U_6=\varepsilon\,\,,\text{and} \, U_2=U_3=U_5=1$.
Moreover, expanding the equations of motion near  $y=0$  gives $U_1(x,0)=U_2(x,0)$.
The reference metric $\tilde{g}$ to be given by the line element (\ref{eq:ansatzbh}) with
$U_4=0, U_6=\varepsilon$ and $U_1=U_2=U_3=U_5=1$.

In the left  panels of Fig.~\ref{fig:BH}, we show  the typical quadrupolar  rotation results of $U_4$  for large black hole in the  right top panel  and  for small black hole in the  right bettom panel, which correspond  to  $\varepsilon=2.215$ and $\varepsilon=2.5$,  respectively.
 In order to compare with the results of the dipolar different boundary rotations for the large black hole,   in the top right we show the distributions of $U_4$ as a function of  the $y$ coordinate  at  the equatorial plane    $x= 0$ for various
values of $\varepsilon$ with quadrupolar (red lines) and dipolar ( blue lines) rotation, respectively.  Meanwhile, in the bottom right the distributions of $U_4$  for small black hole are shown.
From two plots in  the left panels, we can see that the deformation of the curve of $U_4$  arose  with the increasing $\varepsilon$, and the curves of the  quadrupolar  rotation have also more twists and turns than that with the  dipolar  rotation,  which is similar to the soliton case.

%%%%%%%%%%%%%%%%%%%%%%%%%%%%%%%%%%%%%%%%%%%%%%%%%%%%%%%%%%%%%%%%%%%
\subsection{\label{subsubsec:ent}Entropy}
%%%%%%%%%%%%%%%%%%%%%%%%%%%%%%%%%%%%%%%%%%%%%%%%%%%%%%%%%%%%%%%%%%
\begin{figure}[t]
\centering
  \begin{minipage}[t]{0.49\textwidth}
    \includegraphics[width=\textwidth]{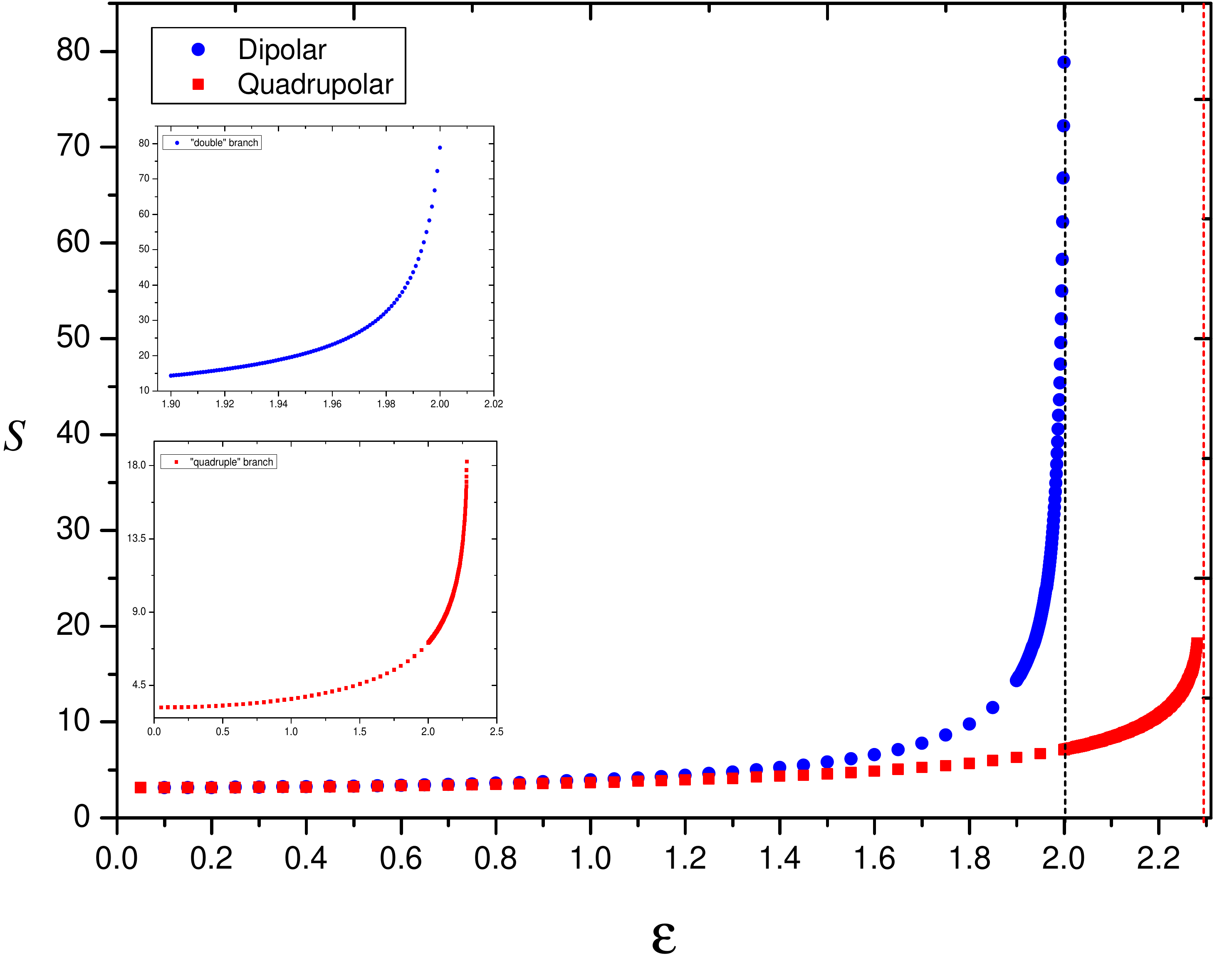}
  \end{minipage}
    \hfill
    \begin{minipage}[t]{0.5\textwidth}
    \includegraphics[width=\textwidth]{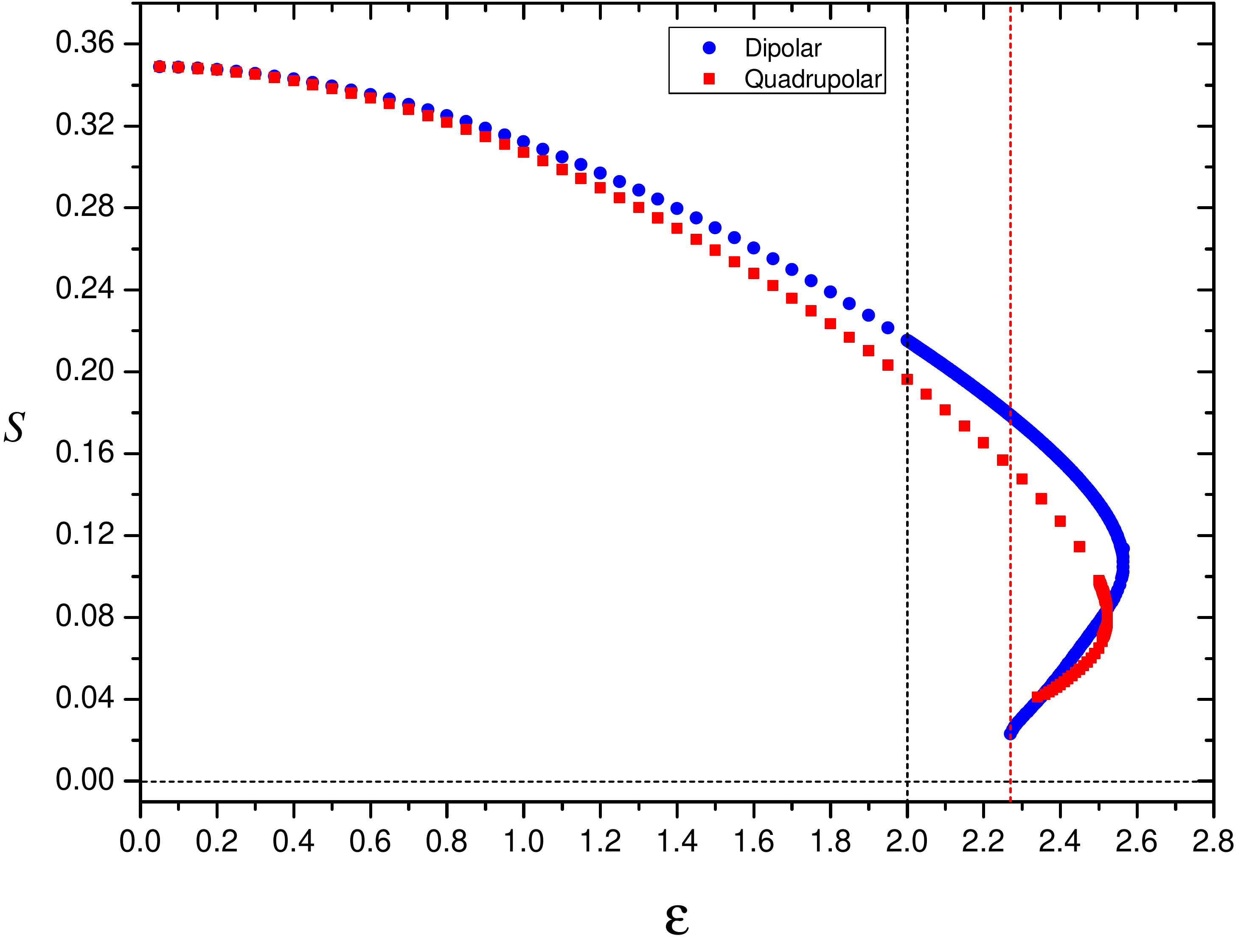}
  \end{minipage}
   \caption{Entropy versus the parameter $\varepsilon$ for the small and large branches of black hole solutions
with a temperature $T=1/\pi$. \textit{Left}: The large  black hole of dipolar (blue line) and quadrupolar (red line) differential rotation with $y_p=1$ are shown. \textit{Right}: The small black hole of dipolar  and quadrupolar  differential rotation with $y_p=1/3$ is shown with blue and red lines, respectively. The horizonal black  dashed gridlines marks the entropy $S=0$.
In both panels the  vertical   black and red  dashed gridlines indicate the  $\varepsilon=2$ and $\varepsilon=2.281$ maximum value, respectively.
}
  \label{fig:entQuadruple}
\end{figure}
%%%%%%%%%%%%%%%%%%%%%%%%%%%%%%%%%%%%%%%%%%%%%%%%%%%
\begin{figure}[t]
\centering
  \begin{minipage}[t]{.7\textwidth}
    \includegraphics[width=\textwidth]{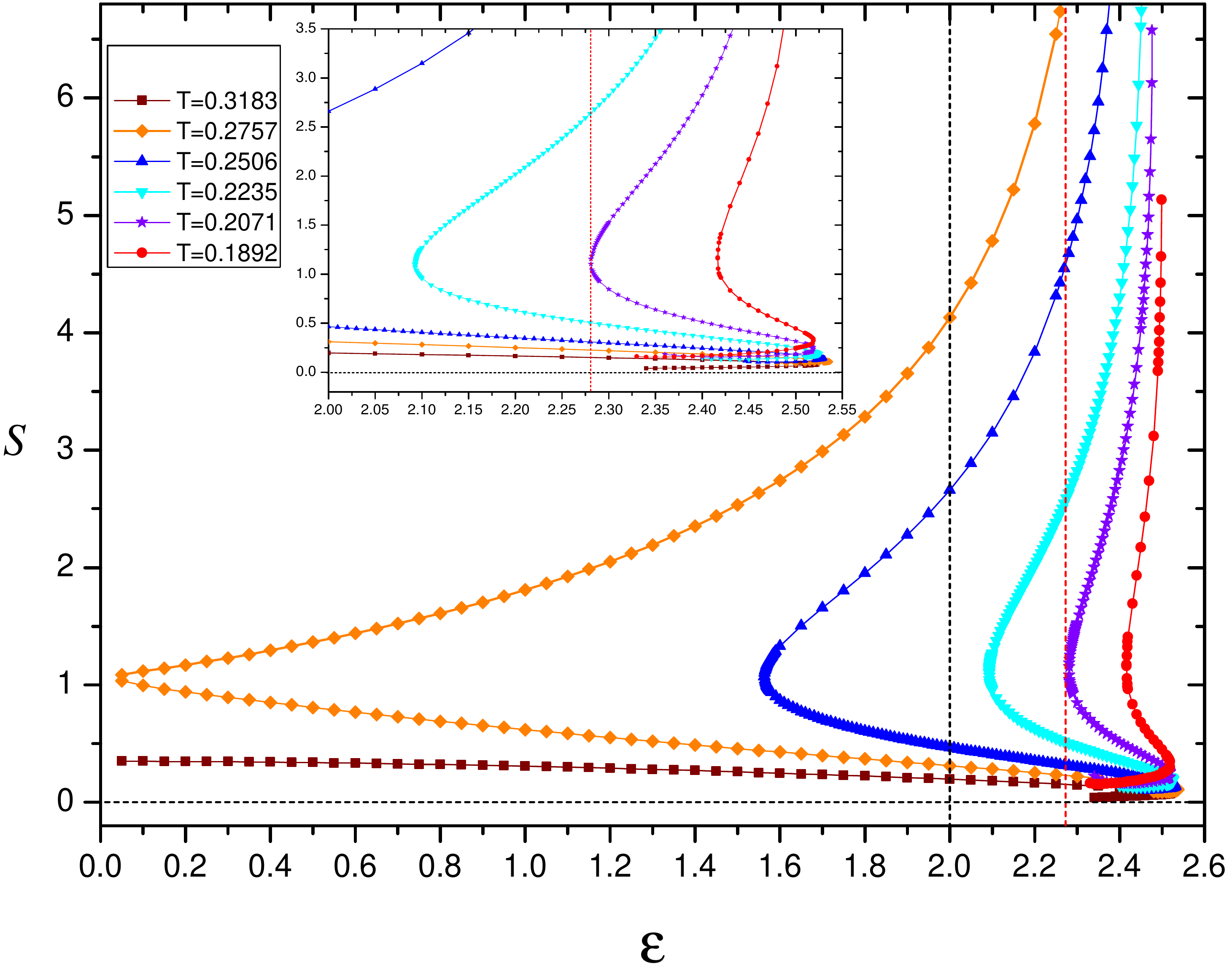}
  \end{minipage}
  \caption{ Entropy versus the parameter $\varepsilon$ for low temperature black holes with $T<T_{c}\simeq0.2757$. The brown squares show the small black holes with $T=1/\pi$, the  vertical   black and red  dashed gridlines indicate the  $\varepsilon=2$ and $\varepsilon=2.281$ maximum value, respectively.}
\label{fig:entropyQuadruple}
\end{figure}
%%%%%%%%%%%%%%%%%%%%%%%%%%%%%%%%%%%%%%%%%%%%%%%%%%%%%%%%%%%%%%%%%%%
In this section we discuss the  entropy of deforming  black holes with quadrupolar differential rotation, which    is proportional to the area of  event horizon and   given by                                                                                                                                                                                                                                                                                                                                                                                                                                                                                                                                                                                                                                                                                                                                                                                                                                                                                                                                                                                                                                                                                                                                                                                                                                                                                                                                                                                                                                                                                                                                                                                                                                                                                                                                                                                                                                                                                                                                                                                                                                                                                                                                                                                                                                                                                                                                                                                                                                                                                                                                                            \begin{equation}
S=\frac{A}{4 G_N}=\frac{2\pi\,y_p^2\,L^2}{G_N}\int_0^1\mathrm{d}x\frac{1-x^2}{\sqrt{2-x^2}} \sqrt{U_3(x,0)U_5(x,0)}\,.
\end{equation}
%%%%%%%%%%%%%%%%%%%%%%%%%%%%%%%%%%%%%%%%%%%%%%%%%%%%%%%%%%%%%%%%%%
In Fig.~\ref{fig:entQuadruple}, we show  entropy versus the parameter $\varepsilon$ for the small and large branches of black hole solutions
with the temperature $T=1/\pi$. The large  black hole with $y_p=1$ is shown in the left panel, where the red  and blue lines correspond to   black holes    with quadrupolar  and dipolar differential rotation, respectively.  The vertical  red  dashed gridlines indicate the  $\varepsilon=2.281$ maximum value beyond which one cannot find axially symmetric black hole solutions with quadrupolar differential rotation.   Meanwhile, the vertical  black  dashed gridlines indicates the  $\varepsilon=2$ maximum value of solutions with dipolar differential rotation.
For the large black holes with quadrupolar differential rotation,  the entropy  always increases with the increasing of $\varepsilon$, which is similar as the case of dipolar  differential rotation.
For the small black hole  with  $y_p=1/3$ in the right panel,  the red  and blue lines correspond to solutions  with quadrupolar and dipolar differential rotation, respectively.
With the increasing of $\varepsilon$, the entropy of  black hole with quadrupolar differential rotation decreases firstly and then  reaches a maximum point of of $\varepsilon$.
Further decreasing $\varepsilon$, one can find another branch of small black holes in  which the   entropy continues to decrease.

 Comparing  with the results of dipolar differential rotation, we find  that
 the norm of Killing vector  $\partial_t$ in Eq. (\ref{eq:timelike}) becomes spacelike for certain regions of $\theta$ when $\varepsilon\in(2,2.281)$,
however, black holes with quadrupolar differential rotation  do not develop hair due to superradiance, which  was different from the case of  dipolar rotation.
At the fixed temperature $T>T_{c}\equiv\sqrt{3}/(2\pi)\approx 0.2757$, the entropies of the large and small black hole  have the  similar behaviors as those in Fig.~\ref{fig:entQuadruple}.
In order to study the solutions at $T<T_{c}\equiv\sqrt{3}/(2\pi)$, we  fix the value of the temperature
with $\delta<1$. When $\delta\rightarrow 0$, the deformed black holes can have
a minimum temperature arbitrarily close to zero
temperature.

In Fig. \ref{fig:entropyQuadruple}, the entropy against the deforming  parameter for low temperature black holes with
 $T<T_{c}\equiv\sqrt{3}/(2\pi)$ is shown. The  vertical   black and red  dashed gridlines indicate the  $\varepsilon=2$ and $\varepsilon=2.281$, respectively.
and  the brown squares  represent the  small black holes with $T=1/\pi$,  which has been discussed in Fig. \ref {fig:entQuadruple}.
 The orange disks show black holes with the critical temerature $T_{c}\simeq0.2757$, and  two branches of  black holes at critical temperature begin to
connect and  form a curve. As the temperature continue to decrease,  the
turning point corresponds to higher values of $\varepsilon$.

%%%%%%%%%%%%%%%%%%%%%%%%%%%%%%%%%%%%%%%%%%%%%%%%%%%%%%%%%%%%%%%%%%%%
\subsection{\label{subsubsec:theh}Horizon geometry}
%%%%%%%%%%%%%%%%%%%%%%%%%%%%%%%%%%%%%%%%%%%%%%%%%%%%%%%%%%%%%%%%%%%
\begin{figure}[t]
\centering
  \begin{minipage}[t]{.48\textwidth}
    \includegraphics[width=\textwidth]{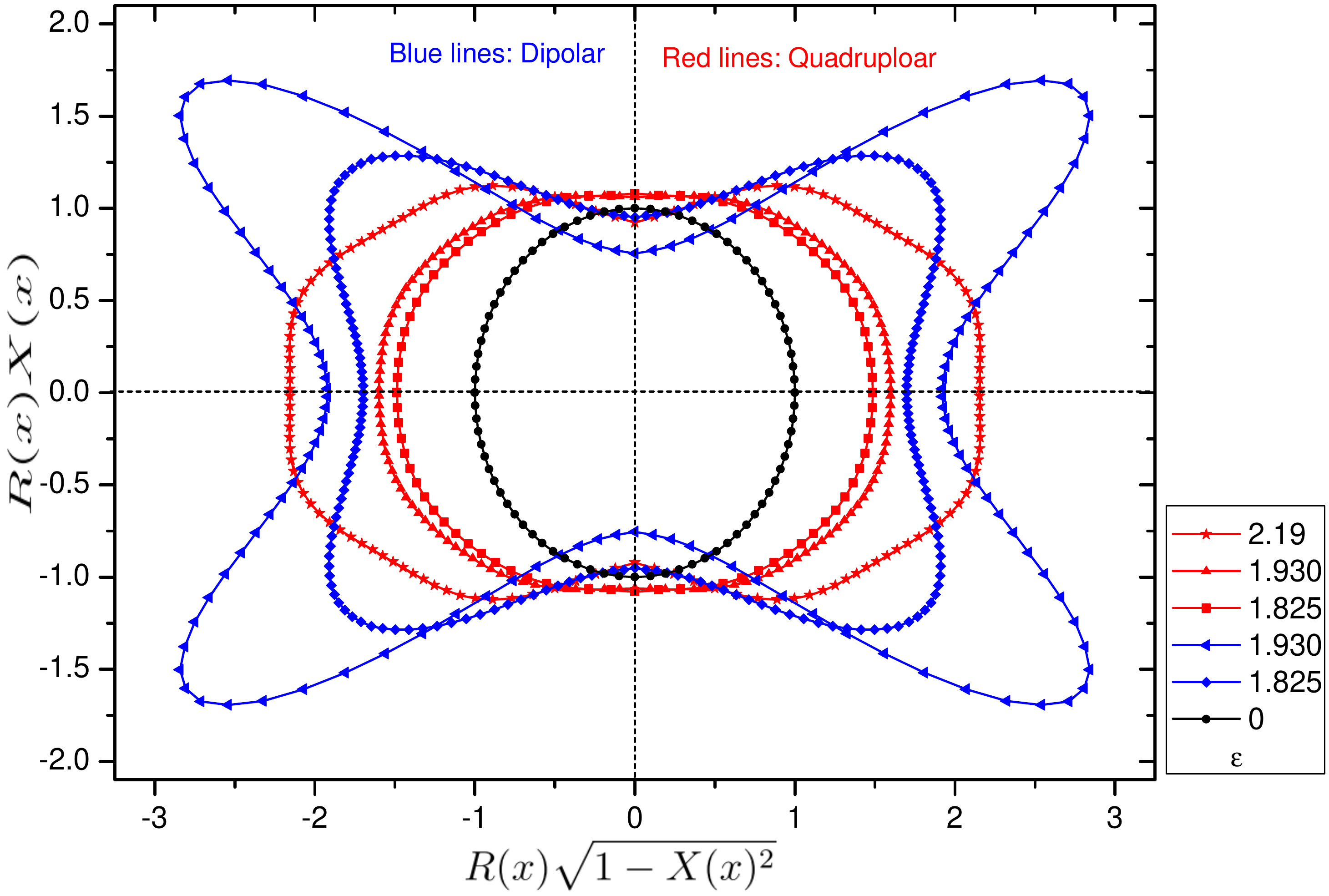}
  \end{minipage}
  \hfill
    \begin{minipage}[t]{.48\textwidth}
    \includegraphics[width=\textwidth]{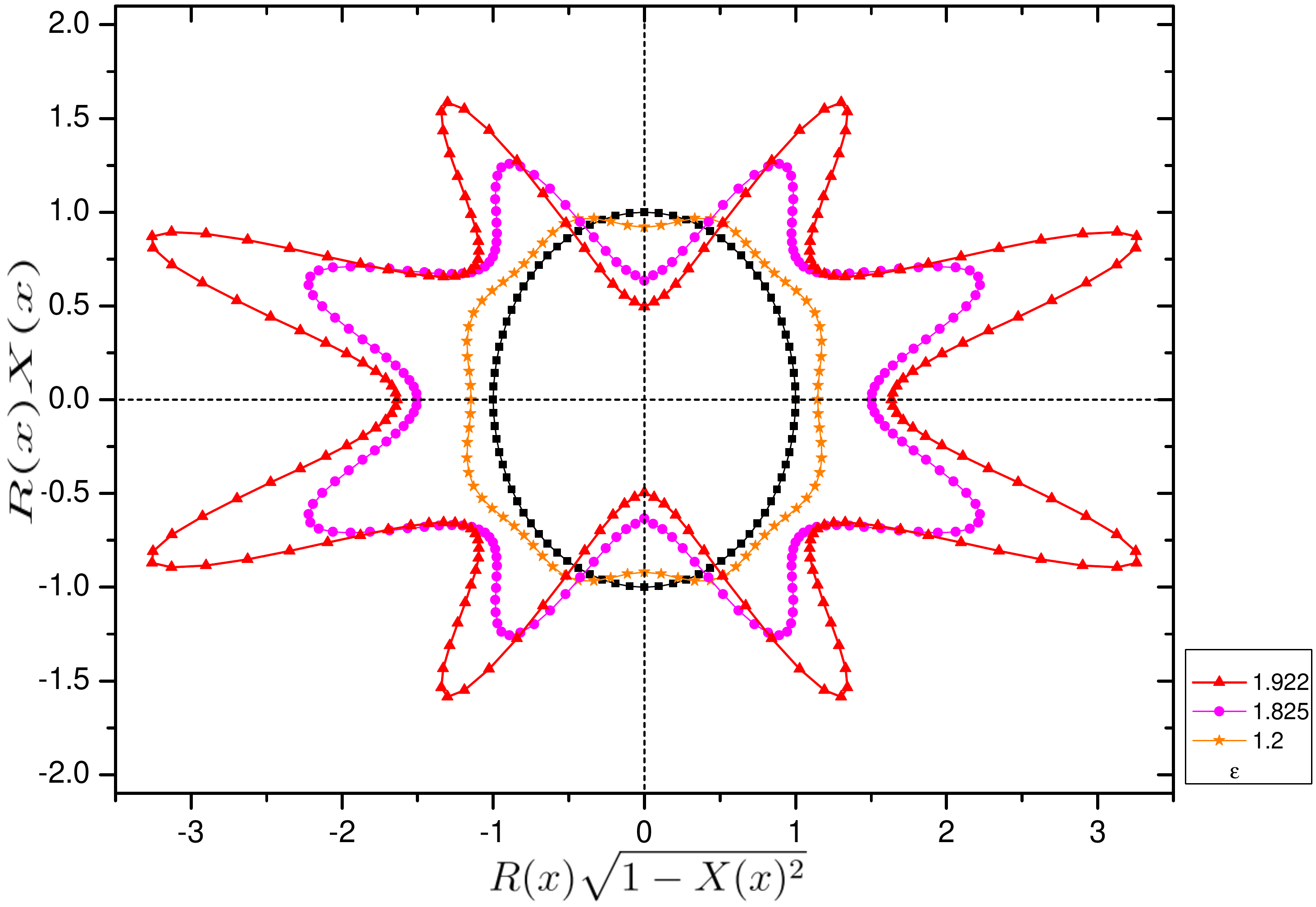}
  \end{minipage}
  \caption{\textit{Left}: Hyperbolic embedding of the cross section of the black hole horizons at the temperature $T=1/\pi$.
 The blue and red lines correspond  to  dipolar and quadrupolar rotation, respectively.
\textit{Right}: Hyperbolic embedding of the cross section of the black hole horizons at the temperature $T=2/\pi$. In both panels the black dashed lines represent the  Schwarzschild-AdS black hole at the temperature $T=1/\pi$. }
  \label{fig:hype}
\end{figure}

%%%%%%%%%%%%%%%%%%%%%%%%%%%%%%%%%%%%%%%%%%%%%%%%%%%%%%%%%%%%%%%%%%%%
Though  the numerical results of the metric (\ref{eq:ansatzbh}) are obtained in last section,  one typically has little information about the real
geometry features of even horizons of deforming black hole in coordinate
space. In order to see how the deforming boundary affects the geometric features  of the event horizon, we can  investigate the geometry of a two-dimensional surface in a curved space by using an isometric embedding in the three-dimensional space \cite{Flamm,Smarr,Friedman,Rosen,Goenner}, which  has been  introduced to study the horizon with dipolar rotation embedding in hyperbolic space\cite{Gibbons:2009qe,Markeviciute:2017jcp}.
In the polar coordinates, the metric of  hyperbolic three-dimensional space $\mathbb{H}^3$ is given by
\begin{equation}\label{eq:hyper3}
\mathrm{d}s^2_{\mathbb{H}^3}=
\frac{\mathrm{d}R^2}{1+R^2/\tilde{\ell}^2}+R^2\left[\frac{\mathrm{d}X^2}{1-X^2}+
(1-X^2)\,\mathrm{d}\phi^2\right],
\end{equation}
where $\tilde{\ell}$ is the radius of the hyperbolic space, and the induced metric on the horizon of the black hole with the metric (\ref{eq:ansatzbh}) is given by
\begin{equation}
\mathrm{d}s^2_H=L^2 \left[\frac{4 y_p^2 U_3(x,0)}{2-x^2}\,\mathrm{d}x^2+y_p^2(1-x^2)^2U_5(x,0)\,\mathrm{d}\phi^2\right],
\label{eq:horizon}
\end{equation}
 and  with the pull back of  line element (\ref{eq:hyper3})  on the induced metric (\ref{eq:horizon}), one can obtain  a embedding of  two-dimensional line element, which is given by a parametric curve $\{R(x),X(x)\}$.

The numerical results of hyperbolic embedding of the cross section of the event horizons for several values of $\varepsilon$ are presented in Fig.~\ref{fig:hype}. In order to compare with the results of dipolar rotation in \cite{Markeviciute:2017jcp}, we also adopt the same parameter $\tilde{\ell}=0.73$.
In the left panel,  we fix the black hole temperature to be $T=1/\pi$, the black line represents the   curve  of Schwarzschild-AdS black hole with  $\varepsilon=0$. The blue and red lines correspond  to  dipolar and quadrupolar rotation, respectively. As one increases $\varepsilon$,
the horizon
cross section begin to deform and has the four arms of the horizon
cross section, which are taken further apart and form the quadrupolar structure.
Comparing with the dipolar rotation, the deformation of quadrupolar rotation is small. In the right panel, at the temperature $T=2/\pi$, we  recalculate the embedding of the cross section and  obtain the larger deformation of quadrupolar rotation.

\begin{figure}[t]
\centering
  \begin{minipage}[t]{.49\textwidth}
    \includegraphics[width=\textwidth]{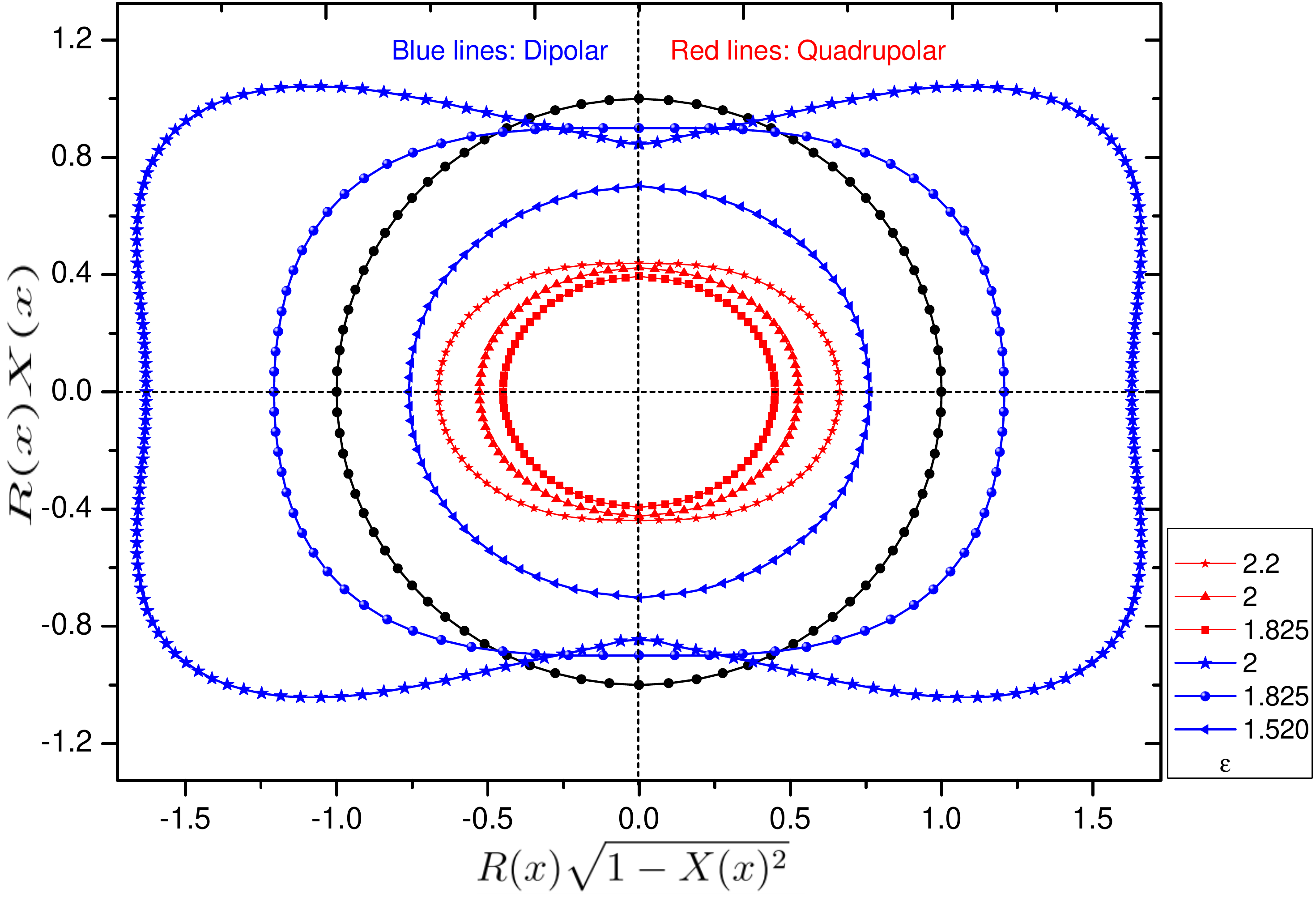}
  \end{minipage}
    \begin{minipage}[t]{.49\textwidth}
    \includegraphics[width=\textwidth]{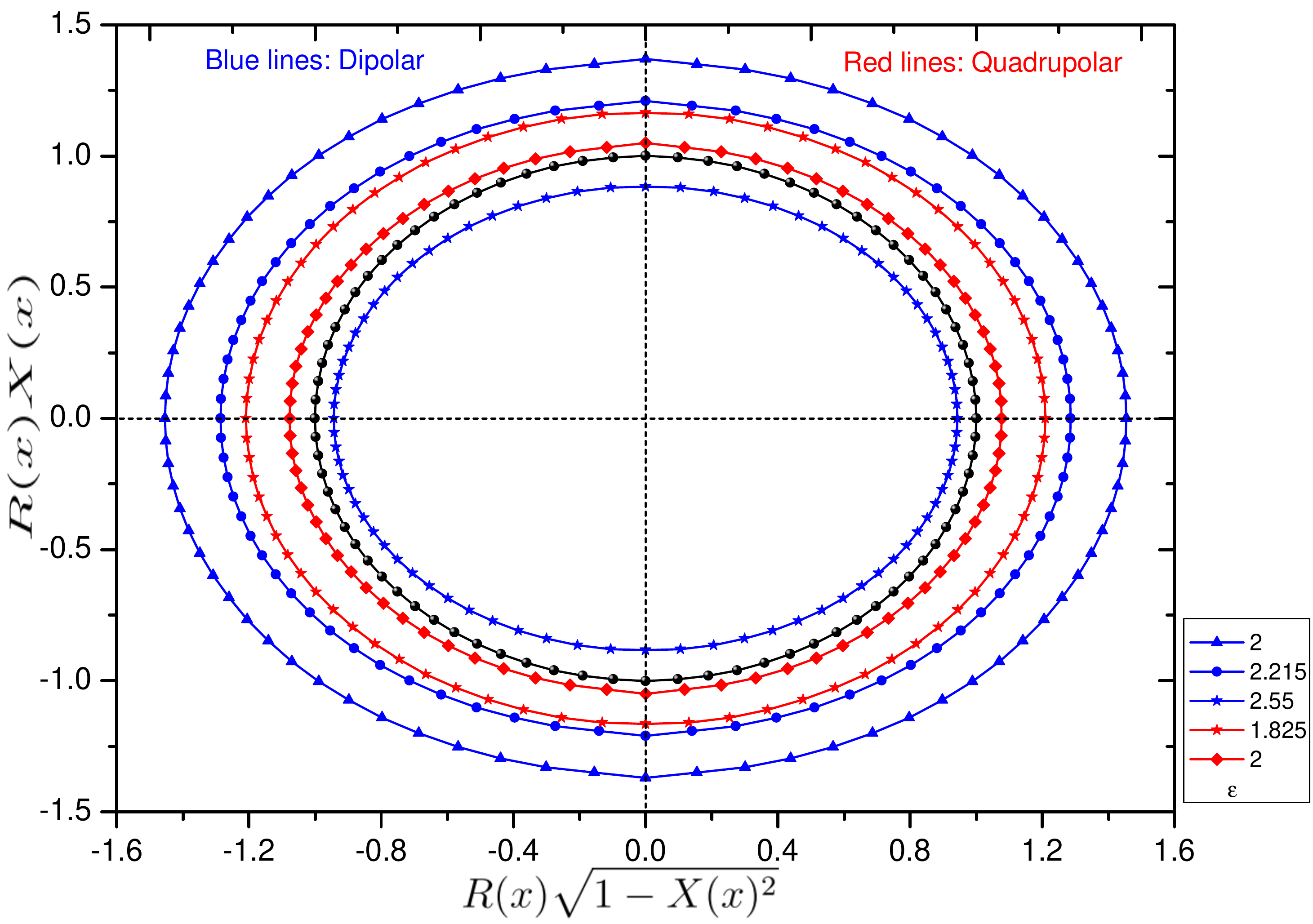}
  \end{minipage}
  \caption{Hyperbolic embedding of the cross section of the black hole solution horizons with low temperature $T=0.2506< T_c$. The black circle  represents the black hole with $\varepsilon=0$ at the temperature $T = 1/\pi$. \textit{Left}: Embeddings for the large black hole solutions.  \textit{Right}: Embeddings for the small black hole branches. In both panels the blue lines represent dipolar rotation as well as the red lines represent quadrupolar rotation.}
  \label{fig:hypeLOW}
\end{figure}

 In ~Fig.~\ref{fig:hypeLOW}, we set a low temperature $T=0.2506$, and find that the deformation of quadrupolar rotation is  smaller with the decrease of temperature. Moreover, the geometry of the horizon
cross section shrink to the interior. Comparing with the curve of  the large black hole in the left panel, the small black hole have  nearly circular curve.
At the same temperature, the curve of  quadrupolar rotation is closer to the interior than that of dipolar rotation.

%%%%%%%%%%%%%%%%%%%%%%%%%%%%%%%%%%%%%%%%%%%%%%%%%%%%%%%%%%%%%%%%%%

\subsection{\label{subsubsec:QNM}Quasinormal modes}
In this subsection we will discuss  the linear stability of deforming black hole with quadrupolar  rotation by studying the quasinormal modes.
%%%%%%%%%%%%%%%%%%%%%%%%%%%%%%%%%%%%%%%%%%%%%%%%%%%%%%%%%%%%%%%%%%
\begin{figure}[t]
\centering
    \begin{minipage}[t]{0.7\textwidth}
    \includegraphics[width=\textwidth]{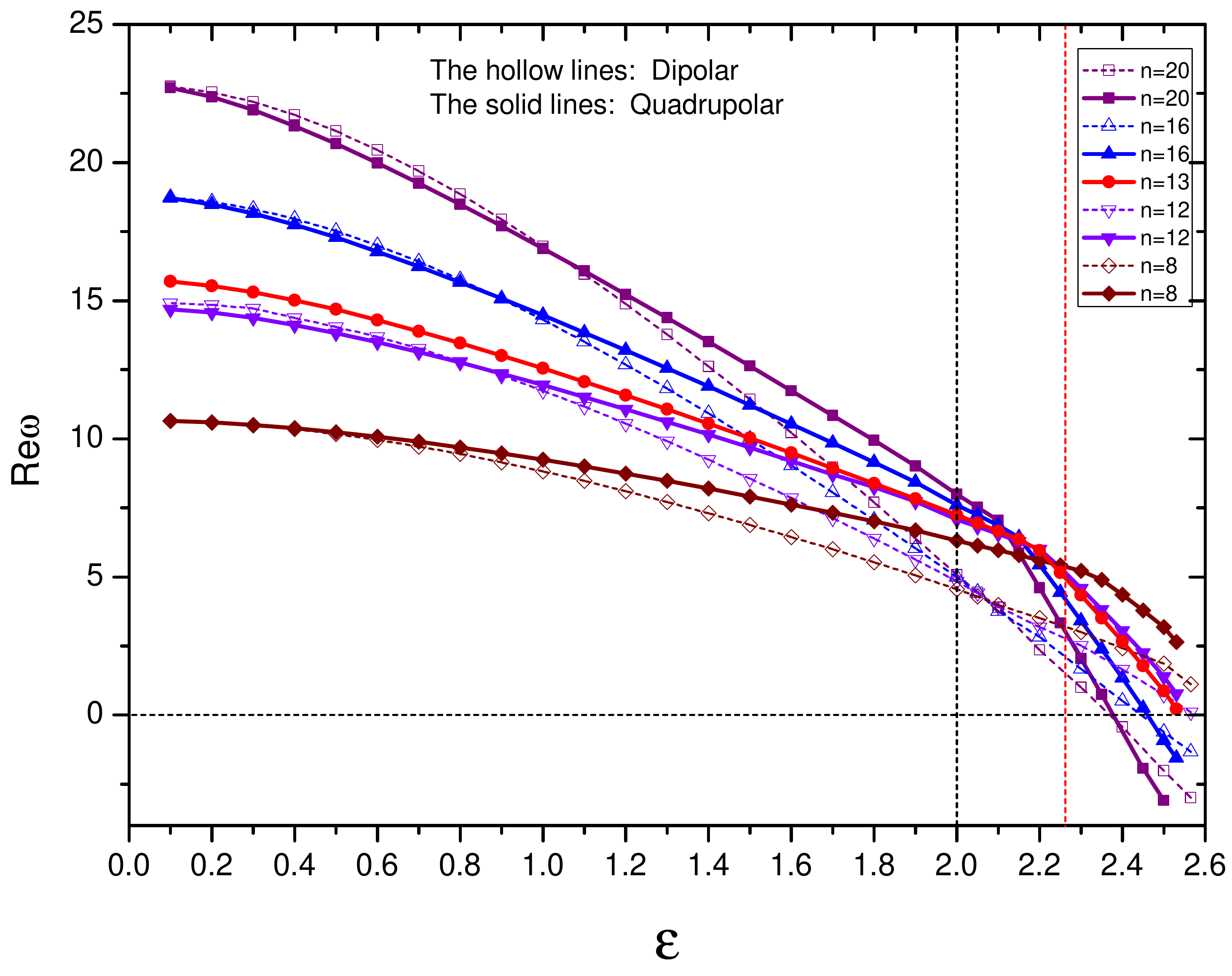}
  \end{minipage}
  \caption{The  real part of frequencies $\omega$  as a function of the deforming parameter $\varepsilon$. The solid lines and dashed hollow lines   stand for the quadrupolar  and dipolar  solutions, respectively.
The  vertical   black and red  dashed gridlines indicate the  $\varepsilon=2$ and $\varepsilon=2.281$, respectively.}   \label{fig:BHStability}
\end{figure}
With the  ansatze of the black hole metric ~(\ref{eq:ansatzbh}), the scalar
field imposed regularity in ingoing Eddington-Finkelstein coordinates \cite{Horowitz:1999jd,Berti:2009kk} could be
decomposed into
\begin{equation}
\chi (t,x,y,\chi)=e^{i(-\omega t+n\chi)}y^{-i\frac{2 \omega y_p}{1 + 3 y_p^2}} (1-y^2)^3 (1-x^2)^{|n|}\psi(x,y)\,,
\end{equation}
where the powers of $x$ and $y$ were chosen to make function $\psi(x,y)$ to be regular at the origin.
At $y=0$ and  $x=\pm1$,  we require that the function $\psi(x,y)$ approach the homogeneous
solution with Neumann boundary conditions.
In addition, the boundary conditions at $y=1$
are given by
\begin{equation}
\partial_y \psi(x,1)=\frac{2 i\,y_p \omega}{(1+3 y_p^2)}\psi(x,1).
\end{equation}

In the Fig~\ref{fig:BHStability},  for a small black black hole  at  $T=1/\pi$,
we plot  the  real part of frequencies $\omega$  as a function of the deforming parameter $\varepsilon$   for the corresponding  values of  $n$, represented by solid lines. In addition,  we also plot the curve of QNMs  with dipolar  rotation studied in \cite{Markeviciute:2017jcp}, represented by dashed hollow lines.
 The  vertical   black and red  dashed gridlines indicate the  $\varepsilon=2$ and $\varepsilon=2.281$, respectively, and  the  horizonal   black  dashed gridlines  shows where  Re\;$\omega$ = 0.
From the figure, we can see that  frequencies  Re\;$\omega$ with $n\leq13$ are always positive values in the spectrum of perturbations, while the frequency begin to be negative at a specific value of $\varepsilon$ when $n\geq n_c=14$.  The characteristic of Re\;$\omega$ against the boundary rotation parameter $\varepsilon$ is  similar to that of  soliton solutions. For the first
unstable mode at $m = 14$,
one can expect some branches of black hole with scalar hair $\Phi$ condensation can be found.

%%%%%%%%%%%%%%%%%%%%%%%%%%%%%%%%%%%%%%%%%%%%%%%%%%%%%%%%%%%%%%%%%%%%%%%

%\newpage
\section{\label{sec:con}Conclusions}
%%%%%%%%%%%%%%%%%%%%%%%%%%%%%%%%%%%%%%%%%%%%%%%%%%%%%%%%%%%%%%%%%%
In this paper,  we
 analyzed  the conformal boundary of four dimensional static asymptotically AdS solutions in Einstein gravity  and numerically constructed the  solutions of
  compact objects with even multipolar differential rotation boundary, including solitons and black holes.
Comparing with the dipolar differential rotation  solutions in  \cite{Markeviciute:2017jcp},  we found
that for high temperature black holes with $T > T_c \simeq 0.2757$,  the norm of Killing vector  $\partial_t$  becomes spacelike for certain regions of $\theta$ when $\varepsilon\in(2,2.281)$,
however, solitons and black holes with quadrupolar differential rotation  do not develop hair due to superradiance, which  was different from the case of  dipolar rotation. For the large black holes of the high temperature,  we did not find any solutions which could cross $\varepsilon\in 2.281$.
 Furthermore,  with the isometric embedding of horizon, it is clearly seen that  black hole horizon is deformed into  four hourglass shapes.  In addition,  we also study the numerical solutions of the entropies of the large and small black hole, which have the similar behaviors
as that of dipolar differential rotation.   By studying the quasinormal modes,  we discussed  the linear stability of deforming  solitons and  black holes with quadrupolar  rotation, respectively, and found that for some branches of  solution with scalar hair $\Phi$ condensation,
the minimal azimuthal harmonic index $n_c=14$  of the  quadrupolar  differential
rotation is larger than  $n_c=13$  of the  dipolar  differential
rotation.

It is interesting to find that
even though  the norm of Killing vector  $\partial_t$  becomes spacelike for certain regions of $\theta$ when $\varepsilon>2$,
solitons and black holes with quadrupolar differential rotation still exist and  do not develop hair due to superradiance.
We also check the numerical  solutions with hexapolar differential rotation, which show very similar results to  quadrupolar differential rotation.
There exists the solution of hexapolar differential rotation when $\varepsilon>2$.

 There are several  interesting extensions of our work.
First, we have studied  the deforming  black holes with even multipolar differential rotation boundary,  next, we will investigate  black holes with odd multipolar differential rotation boundary, which has the symmetric rotation profile with respect to
reflections on the equatorial plane.  There exists a question whether  total angular momentum of black hole   is  non-zero.
The second  extension of our study is to   consider the action of Einstein-Maxwell gravity in AdS spacetime and construct the deforming charged black holes.
Due to  the existence of charges, at the same temperature, one finds that there are three branches of solutions and the phase diagram of solutions is  more intricate than  that  without charges.
Finally, we are planning to  extend the study of   the deforming  black holes  to the  five-dimensional  solutions  in future work.

\section*{Acknowledgement}
We would like to thank  Yu-Xiao Liu   for  helpful discussion. Some
computations were performed on the   Shared Memory system at  Institute of Computational Physics and Complex Systems in Lanzhou University. This work was supported by the Fundamental Research Funds for the Central Universities (Grants No. lzujbky-2017-182).

%%%%%%%%%%%%%%%%%%%%%%%%%%%%%%%%%%%%%%%%%%%%%%%%%%%%%%%%%%%%%%%%%%%%%%%
%%%%%%%%%%%%%%%%%%%%%%%%%%%%%%%%%%%%%%%%%%%%%%%%%%%%%%%%%%%%%%%%%%%%%%%
\nocite{*}
\bibliographystyle{JHEP}
\bibliography{hairybh}

\end{document}